\documentclass[a4paper,10pt,twocolumn,nofootinbib]{revtex4}

\usepackage{amsmath}
\usepackage{amssymb}
\usepackage[utf8]{inputenc}
\usepackage{tikz}
\usepackage[mathscr]{euscript}
\usepackage{hyperref}
\usepackage{bm}
\DeclareMathAlphabet{\mathsf}{OT1}{\sfdefault}{m}{n}
\SetMathAlphabet{\mathsf}{bold}{OT1}{\sfdefault}{b}{n}

\newcommand\Real{{\mathbb R}}  
\newcommand\union{\bigcup}     
\newcommand\inter{\bigcap}     
\newcommand\Set[1]{{\left\{#1\right\}}} 

\newcommand\qquadtext[1]{\qquad\textup{#1}\qquad}
\newcommand\qquadand{\qquadtext{and}}
\newcommand\quadtext[1]{\quad\textup{#1}\quad}
\newcommand\quadand{\quadtext{and}}

\newcommand\eventZero{\mathbf{0}}
\newcommand\mForm[1]{\boldsymbol{\mathsf{#1}}} 

\newcommand\Hform{\mForm{H}} 
\newcommand\Fform{\mForm{F}} 
\newcommand\Jform{\mForm{J}} 
\newcommand\Bform{\mForm{B}} 
\newcommand\Eform{\mForm{E}} 

\newcommand\Apotl{\mForm{A}} 

\newcommand\Charge{Q}        
\newcommand\manifold[1]{\mathscr{#1}} 

\newcommand\manM{\manifold{M}} 
\newcommand\manN{\manifold{N}} 
\newcommand\manU{\manifold{U}} 
\newcommand\manS{\manifold{S}} 
\newcommand\manC{\manifold{C}} 
\newcommand\manSigma{{\Sigma}} 

\newcommand\de{d}               
\newcommand\hstar{\star}        
\newcommand\boundary{\partial}  

\newcommand{\XDOI}[1]{\href{http://dx.doi.org/#1}{doi:#1}}

\usetikzlibrary{decorations.markings}
\tikzset{->-/.style={decoration={
  markings,
  mark=at position #1 with {\arrow{>}}},postaction={decorate}}}
\tikzset{-<-/.style={decoration={
  markings,
  mark=at position #1 with {\arrow{<}}},postaction={decorate}}}

\newcommand\emVec[1]{\textbf{#1}}
\newcommand\HMag{\emVec{H}}  
\newcommand\BMag{\emVec{B}}  
\newcommand\Elec{\emVec{E}}  
\newcommand\Dlec{\emVec{D}}  

\newcommand\emDH{{$\Dlec$, $\HMag$}}  
\newcommand\emEB{{$\Elec$, $\BMag$}}  

\newcommand\heavi{{\textup{hv}}}
\newcommand\subHP{\textup{g}} 

\newcommand\Jbdd{{\Jform_{\textup{b}}}}  
\newcommand\Jfree{\Jform}                
\newcommand\JFree{{\Jform_{\textup{f}}}} 
\newcommand\Jtot{{\Jform_{\textup{t}}}}  

\newcommand\deRahm{{\textup{dR}}}

\newcommand\RegI{{\textup{I}}}          
\newcommand\RegII{{\textup{I\!I}}}      
\newcommand\RegIII{{\textup{I\!I\!I}}}  
\newcommand\RegIV{{\textup{I\!V}}}      

\newcommand\eqstrut{\rule[-0.5\baselineskip]{0pt}{12pt}} 

\definecolor{XRED}{rgb}{0.70, 0.01, 0.01}
\definecolor{XBLUE}{rgb}{0.01, 0.01, 0.70}
\newcommand\UPDATED[1]{#1}
\newcommand\UPDATES[1]{#1}

\begin{document}
\title{Evaporating black-holes, wormholes, and vacuum polarisation: \\
    must they always conserve charge?}

\author{Jonathan Gratus}
\homepage[]{https://orcid.org/0000-0003-1597-6084}
\email[\hphantom{.}~]{j.gratus@lancaster.ac.uk}
\author{Paul Kinsler}
\homepage[]{https://orcid.org/0000-0001-5744-8146}
\email[\hphantom{.}~]{Dr.Paul.Kinsler@physics.org}
\affiliation{
  Department of Physics, 
  Lancaster University,
  Lancaster LA1 4YB, 
  United Kingdom,
}
\affiliation{
The Cockcroft Institute, 
Sci-Tech Daresbury,
Daresbury WA4 4AD, 
United Kingdom.
}
\author{Martin W. McCall}
\homepage[]{https://orcid.org/0000-0003-0643-7169}
\email[\hphantom{.}~]{m.mccall@imperial.ac.uk}
\affiliation{
  Department of Physics, 
  Imperial College London,
  Prince Consort Road,
  London SW7 2AZ, 
  United Kingdom.
}

\begin{abstract}

A careful examination of the fundamentals of electromagnetic theory
 shows that due to the underlying mathematical assumptions 
 required for Stokes' Theorem, 
 \UPDATES{global} charge conservation cannot be guaranteed
 in topologically non-trivial spacetimes.
However, 
 in order to break the charge conservation mechanism 
 we must also allow the electromagnetic excitation fields {\emDH}
 to possess a gauge freedom, 
 just as the electromagnetic scalar and vector potentials
 $\varphi$ and $\emVec{A}$ do.
This has implications for the treatment of electromagnetism
 in spacetimes where black holes both form and then evaporate, 
 as well as extending the possibilities for treating vacuum polarisation.
Using this gauge freedom of {\emDH}
 we also propose an alternative to the accepted notion
 that a charge passing through a wormhole necessarily
 leads to an additional (effective) charge on the wormhole's
 mouth.

\end{abstract}




\date{\today}

\maketitle

{\small There is a popular summary at the end: \ref{S-popular}.}

\section{Introduction}
\label{ch_INTRO}

It is not only a well established,
 but an extremely useful consequence of Maxwell's equations,
 that charge is conserved
 \cite{Jackson-ClassicalED}.
However,
 this principle relies on some assumptions,
 in particular those about the topology of the underlying spacetime, 
 which are required for Stokes' Theorem to hold.
Here
 we describe how to challenge the status of
 \UPDATES{global charge conservation,
 whilst still keeping local charge conservation intact.
We do this by} investigating the interaction of electromagnetic theory
 and the spacetime it inhabits, 
 and go on to discuss the potential consequences of such a scenario.

As well as
 considering topologically non-trivial spacetimes, 
 we also
 no longer demand that the excitation fields {\emDH}
 are directly measurable.
\UPDATED{This relaxation means that
 the excitation fields {\emDH} are now allowed a gauge freedom
 analogous 
  to that of the scalar and vector electromagnetic potentials
 $\varphi$ and $\emVec{A}$.
This gauge freedom for {\emDH} is given by}
~
\begin{align}
&
  \Dlec 
\rightarrow 
  \Dlec + \nabla\times \emVec{A}_{\subHP}
,
\quad
&
  \HMag 
\rightarrow 
  \HMag + \dot{\emVec{A}}_{\subHP} + \nabla \varphi_{\subHP}
,
\label{eqn-vectorgauge}
\end{align}
   where $\varphi_{\subHP}$
   and $\emVec{A}_{\subHP}$ are the new gauge terms,
 which vanish when inserted into Maxwell's equations.
Note that there are already long-standing debates
 about whether -- 
 or how --
 any measurement of the excitation fields might be done
 (see e.g. \cite{Heras-2011ajp,Gratus-KM-2019ejp-dhfield}
 and references therein).
Unlike the case for {\emEB},
 there is no native Lorentz force-like equation
 for magnetic monopoles dependent on {\emDH},
 although proposals -- 
 based on the assumption that monoples indeed exist --
 have been discussed \cite{Rindler-1989ajp}.
Neither is there an analogous scheme for measuring {\emEB}
 inside a disk by using the Aharonov-Bohm effect
 \cite{Ehrenberg-S-1949prsb,Aharonov-B-1959pr,Matteucci-IB-2003fp},
 a method particularly useful inside a
   medium where collisions may prevent a point charge obeying the
   Lorentz force equation.
This double lack means that 
 whenever making claims about the 
 measurability of {\emDH},
 one has to make assumptions about their nature,
 for example that it is linearly and locally related
 to {\emEB}, 
 such as in the traditional model of the vacuum.
Such assumptions act to fix any gauge for {\emDH},
 so that one can measure the
 remaining parameters; 
 but if {\emDH} are taken to be not measurable,
 then the gauge no longer needs to be fixed.

The relaxed assumptions \UPDATED{about topology and gauge}
 are not merely minor technical details,
 since many cosmological scenarios involve
 a non-trivial topology.
Notably,
 black holes have a central singularity that is missing
 from the host spacetime {\cite{Schutz-FCRelativity,MTW}}, 
 and a forming and then fully evaporated black hole
 creates a non-trivial topology, 
 which in concert with allowing 
 a gauge freedom
 for now \textit{non}-measureable {\emDH} fields,
 breaks the usual basis for charge conservation.
We also consider 
 more exotic scenarios,
 such as the existence of a universe containing a wormhole
 (see e.g. \cite{Morris-T-1988ajp}),  
 or a ``biverse''--
 a universe consisting of two asymptotically flat regions
 connected by an Einstein-Rosen bridge.
In particular we test the claim that 
 charges passing through such constructions (wormholes)
 are usually considered to leave it charged
 \cite{Visser-LW,Susskind-2005arXiv,Wheeler-1957ap}.

Topological considerations
 and their influence on the conclusions of Maxwell's theory are not new,
 but our less restrictive treatment of {\emDH} 
 allows us a wider scope than in previous work.
Misner and Wheeler,
 in \cite{Wheeler-1957ap} developed an ambitious programme
 of describing all of classical physics
 (i.e. electromagnetism and gravity) geometrically,
 i.e. \textit{without} including charge at all.
Non-trivial topologies,
 such as spaces with handles,
 were shown to support situations where charge could be interpreted as
 the non-zero flux of field lines,
 which never actually meet,
 over a closed surface containing the mouth of a wormhole. 
Baez and Muniain \cite{BaezMunain-GFKG}
 show that certain wormhole geometries are simply connected,
 so that every closed 1-form is exact. 
In this case charge can then be \textit{defined}
 as an appropriate integral of the electric field over a 2-surface.
In another example,  Diemer and Hadley's
 investigation \cite{Diemer-H-1999cqg} has shown that it is possible,
 with careful consideration of orientations,
 to construct {wormhole} spacetimes containing 
 topological magnetic monopoles or topological charges; 
\UPDATED{and 
 Marsh \cite{Marsh-1998jpa} has discussed
 monopoles and gauge field in electromagnetism
 with reference to topology and de Rham's theorems.}


It is important to note
 that our investigation here is entirely 
 distinct from and prior to 
 any cosmic censorship conjecture \cite{Penrose-1999jaa},
 the boundary conditions at a singularity,
 models for handling the event horizon \cite{Price-T-1986prd},
 or other assumptions.
Although an event horizon or other censorship arrangement
 can hide whatever topologically induced effects there might be,
 such issues are
 beyond the scope of our paper,
 which instead focuses on the fundamental issues --
 i.e. the
 prior and {classical} consequences
 of the violation of the prerequisites of Stokes' Theorem
 in spacetimes of non-trivial topology.

In section \ref{ch_EM} we summarise the features of electromagnetism
 relevant to our analysis. 
In section \ref{ch_Singular} we investigate 
 under what circumstances charge conservation no longer holds, 
 and its consequences for the electromagnetic excitation field $\Hform$, 
 which is the differential form version
  of the traditional {\emDH}.
Next, 
 in section \ref{ch_Pol} we describe further consequences, 
 such as
 how a description of bound and free charges necessarily supplants
 a standard approach using constitutive relations based on $\Hform$.
Then, 
 in section \ref{ch_Worm} we see that topological considerations
 mean that $\Hform$ can be defined in a way that has implications
 for the measured charge of wormholes.
\UPDATED{Lastly,
 after some discussions in section \ref{ch_Discussion},
 we summarise our results in section \ref{ch_Conclude}.}

\section{Electromagnetism}
\label{ch_EM}

\subsection{Basics}

Although perhaps the most famous version of Maxwell's equations are
 Heaviside's vector calculus form
 in  {\emEB} and {\emDH},
 here we instead use the language of differential forms
 \cite{Flanders1963,HehlObukhov}, 
 an approach particularly useful when treating electromagnetism
 in a fully spacetime context \cite{McCall-FMB-2011jo,Gratus-KMT-2016njp-stdisp,Cabral-L-2017fp}.
This more compact notation combines the separate time and space behaviour
  into a natively spacetime formulation,
 so that the four vector equations in curl and divergence
 are reduced to two combined Maxwell's equations
 \cite{Flanders1963,HehlObukhov}:
\begin{align}
  {\de} \Fform
=
  0,
\label{CQ_Max_F}
\end{align}
and
\begin{align}
  {\de} \Hform 
=
  \Jfree
.
\label{CQ_Max_HJ}
\end{align}
Here $\Fform, \Hform \in\Gamma \Lambda^2 \manM$ are the
 Maxwell and excitation 2-form fields on spacetime $\manM$,
  $\Jfree\in\Gamma\Lambda^3 \manM$ is the free current density 3-form, 
 \UPDATED{and $\Gamma$ indicates 
 that $\Fform$ and $\Hform$ are 
 smooth global sections of the bundle $\Lambda^2 \manM$.}
Conventionally,
 $\Fform = \Eform \wedge {dt}+\Bform$,
 where $\Eform$ is a 1-form representing the electric field, and $\Bform$ 
 is a 2-form representing the magnetic field.
Taking the exterior derivative
 (${\de}$)
 of \eqref{CQ_Max_HJ}
  leads to the differential form of charge conservation,
 i.e. that $\Jfree$ is closed,
\begin{align}
  {\de} \Jfree 
=
  0
.
\label{CQ_Max_dJ}
\end{align}
As
 equations \eqref{CQ_Max_F} and \eqref{CQ_Max_HJ} are underdetermined,
 they need to be supplemented by a constitutive relation,
 connecting $\Fform$ and $\Hform$.
In general this relation can be arbitrarily complicated,
 but the simplest is the ``Maxwell vacuum''
 where they are related by the Hodge dual ${\hstar}$,
 i.e.
\begin{align}
  \Hform 
=
  {\hstar} \Fform
.
\label{CQ_CR_vac}
\end{align}
It is worth noting,
 however,
 that competing constitutive models exist,
 even for the vacuum.  
Two well known examples are
 the weak field Euler-Heisenberg constitutive relations
 and Bopp-Podolski
 \cite{Bopp-1940ap,Podolsky-1942pr,Gratus-PT-2015jpa} 
 constitutive relations, 
which are respectively
\begin{align}
  \Hform_{\textup{EH}}
&=
  {\hstar} \Fform 
 -
  \frac{8\alpha^2}{45 m^4}
  \left[
    {\hstar} \left( \Fform\wedge{\hstar} \Fform \right)
    {\hstar} \Fform
   +
    7 {\hstar} \left(\Fform\wedge \Fform\right)\, 
    \Fform
  \right]
,
\label{CQ_CR_EH}
\end{align}
and
\begin{align}
  \Hform_{\textup{BP}} 
&=
  {\hstar} \Fform + \ell^2 {\hstar} {\de} {\hstar} {\de} {\hstar} \Fform
,
\label{CQ_CR_BP}
\end{align}
where $\alpha$ is the fine structure constant, $m$ is the mass of
the electron and
length $\ell$ is a small parameter.

However, 
 fixing a constitutive relation where $\Hform$
 has a straightforward relationship to $\Fform$,
 such as those given above,
 is of itself sufficent to enforce charge conservation.
In contrast,
 we consider more general constitutive models,
 and so can investigate wider possibilities.

\begin{figure}
\centering
\resizebox{0.480\columnwidth}{!}{ 
\includegraphics{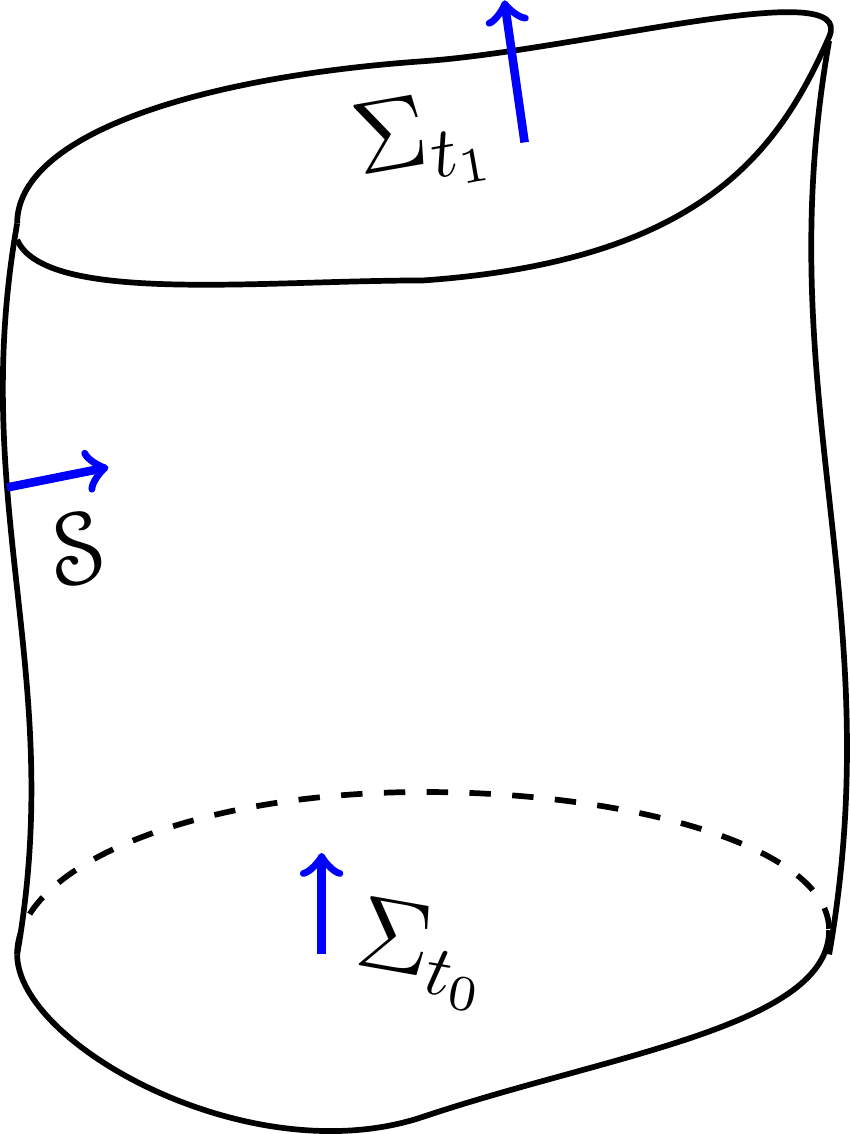}
}
\caption{\UPDATED{A closed 3-surface $\manU$ in spacetime on which to check 
 conservation of charge.
This surface is formed from 
 $\manU = \manS \union \manSigma_{t_0} \union
  \manSigma_{t_1}$,
 with the orientations of $\manS$, $\manSigma_{t_0}$ and $\manSigma_{t_1}$
  given by the blue arrows.
Note that we do not necessarily need to consider
 a 4-volume enclosed by this boundary $\manU$,
 as can be seen by comparing the topology condition
 with the 
 gauge-free condition,
 as discussed in the main text.
}}
\label{fig_consCharge}
\end{figure}

\subsection{Conservation of charge}
\label{ch_CQ}

The starting point for our investigation of topology, 
 charge conservation, 
 and the role of $\Hform$,
 is a closed 3-surface $\manU$
 with no boundary, 
 i.e. ${\boundary} \manU = \emptyset$.
\UPDATED{This surface $\manU$
 is topologically equivalent to the 3-sphere,
 and is depicted on fig. \ref{fig_consCharge}.}
We can write $\manU = \manS \union \manSigma_{t_0}\union \manSigma_{t_1}$
 where $\manSigma_{t_0}$ and $\manSigma_{t_1}$ are bounded regions
 of the space $\manSigma$ at times $t_0$ and $t_1$, 
 and $\manS$ is the boundary of $\manSigma$
 between the times $t_0\le t\le t_1$.  
As shown in fig. \ref{fig_consCharge}, 
 the orientation of $\manSigma_{t_1}$ is outward, 
 while those of $\manSigma_{t_0}$ and $\manS$ are inward.
Charge conservation,
 expressed as
\begin{align}
\int_\manU \Jfree = 0,
\label{CQ_int_U_J}
\end{align}
 can be expressed in this case as
\begin{align}
  \int_{\manSigma_{t_1}} \Jfree
 -
  \int_{\manSigma_{t_0}} \Jfree
 -
  \int_\manS \Jfree 
=~~
  \int_\manU \Jfree 
=~~
  0
,
\label{CQ_int_SSig_J}
\end{align}
which we may interpret as the total charge in $\manSigma$ at time $t_1$
 is given by the total charge in $\manSigma$ at time $t_0$,
 plus any charge that entered $\manSigma$ in the time $t_0\le t\le t_1$.

Irrespective of possible complications associated with the
constitutive relations, charge conservation \eqref{CQ_int_U_J} follows
straightforwardly in either of {two} ways,
  \textit{both} due to Stokes' theorem:

\begin{enumerate}

\item \UPDATED{\textbf{Topology condition:}}
The first proof assumes that $\manU$ is the boundary of a
topologically trivial bounded region of spacetime,
 i.e. $\manU={\boundary}\manN$,
 $\manN \subset \manM$,
 within which $\Jfree$ is defined. 
A \textit{topologically trivial} space is one that can be shrunk to a point
i.e. it is topologically equivalent to a 4-dimensional ball.
Then one has 
\begin{align}
  \int_\manU \Jfree 
=
  \int_{{\boundary} \manN} \Jfree 
=
  \int_\manN {\de} \Jfree
=
  0
,
\label{CQ_pf_dJ}
\end{align}
the last equality arising from \eqref{CQ_Max_dJ}, 
 which we call the ``topology condition''.

\item  \UPDATED{\textbf{Gauge-free condition:}}
The second proof arises from integrating \eqref{CQ_Max_HJ} over $\manU$,
 and presumes that $\Hform$ is a well-defined 2-form field.
We have that
\begin{align}
  \int_\manU \Jfree 
=
  \int_\manU {\de} \Hform 
=
  \int_{{\boundary} \manU} \Hform 
=
  0
,
\label{CQ_pf_dH}
\end{align}
 where the last equality, 
 which we call the ``gauge-free condition'',
 results solely from the fact that $\manU$ is
 closed (i.e. ${\boundary} \manU=\emptyset$), but does \textit{not} require
 that $\manU$ is itself the boundary of a {compact} 
 4-volume\footnote{The fact
   that $\Hform$ is well-defined has been used in invoking $\int_\manU
   {\de} \Hform = \int_{{\boundary} \manU} H$ in \eqref{CQ_pf_dH}.
   Compare integrating ${\de}\theta$ around the unit circle $\manC$ to
   obtain the fallacious result $\oint_\manC {\de} \theta = \oint_{{\boundary}
     \manC}\theta = 0$, since ${\boundary} \manC = \emptyset$.  The
   problem is that $\theta$ is not well defined (and continuous) on
   all of $\manC$.  By defining two submanifolds $\manU_1$ and
   $\manU_2$ such that $\manU_1 \cup \manU_2 = \manC$ with respective
   coordinate patches $\theta$ and $\theta+2\pi$, then a careful
   integration around $\manC$ yields the correct answer of $2\pi$.}.

\end{enumerate}

\subsection{Non-conservation of Charge}
\label{ch_NCQ}

The arguments for conservation of charge presented thus far
 have been mathematically rigorous.
Given this sound basis,
 one may ask,
 why would anyone doubt conservation of charge?
One might note,
 for example,
 the case of black holes,
 where charge
  is one of the few quantities preserved in the no-hair theorem
 \cite{Israel-1968cmp}.
However,
 our need to make assumptions about the nature of $\manU$ or $\Hform$
 in the proofs \eqref{CQ_pf_dJ} and \eqref{CQ_pf_dH},
 when establishing conservation of charge,
 provides us with an opportunity for testing its true basis
 and extent of validity.
Notably,
 to create a charge non-conservation loophole,
 \UPDATED{both \eqref{CQ_pf_dJ} and \eqref{CQ_pf_dH}
 must be violated:
 if either one applies then charge is conserved.}

\begin{figure}
\centering
\resizebox{0.80\columnwidth}{!}{
\includegraphics{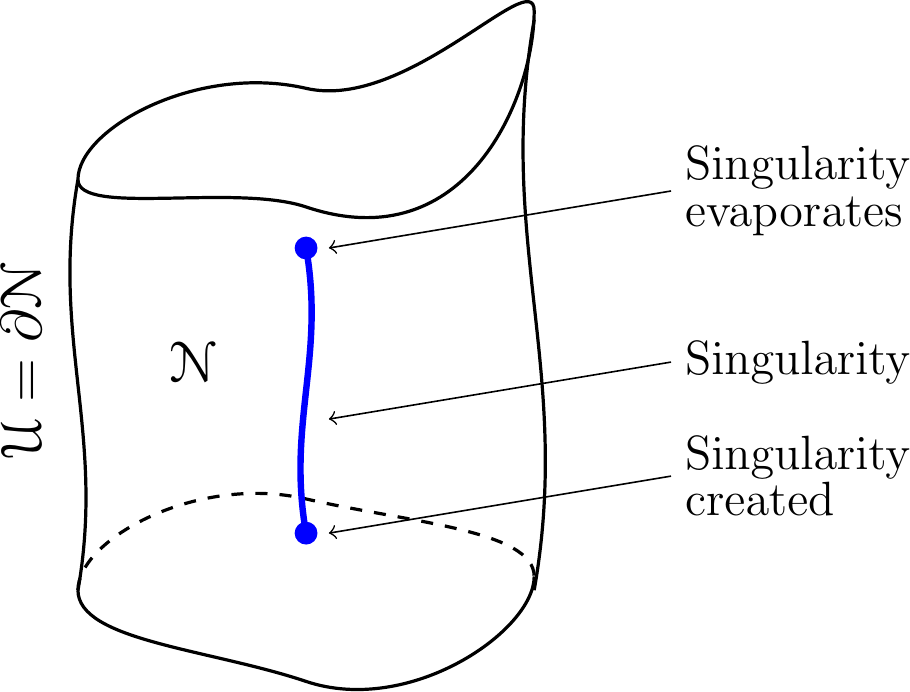}
}
\caption{Here we show a region of spacetime $\manN$
 with a boundary $\manU={\boundary} \manN$ that encloses a singularity
 with a finite duration.
 As a consequence,
  it is topologically non-trivial
  and not globally hyperbolic.
This might occur,
 for example,
 due to the formation and subsequent evaporation of a black hole,
 which would first create and then remove
 a metric singularity in spacetime.
\UPDATED{If the singularity instead existed only for an instant, 
 the blue line drawn here would reduce to a single point, 
 and the figure would then depict the manifold $\manM$
 used in section \ref{ch_Singular}.}
}
\label{fig_singular}
\end{figure}

To break the topology condition \eqref{CQ_pf_dJ},
 it is sufficient that either 
 there is no compact spacetime region $\manN$
 such that $\manU={\boundary} \manN$,
 or that there are events in $\manN$ where $\Jfree$ is undefined.
A test scenario is represented in fig. \ref{fig_singular},
 where a black hole forms
 in an initially unremarkable spacetime, i.e. one that contains
    spatial hypersurfaces that
  are topologically trivial.
On formation this introduces a singularity,
 but then later as the black hole evaporates,
 the singularity also vanishes.
The evaporation step also removes the event horizon,
 thus exposing any effects of the singularity --
 e.g. in charge conservation --
 to the rest of the universe.
%
%
In this case the singularity,
 which exists for a period of time before evaporating
 \cite{Okon-S-2018fp}
 by means of Hawking radiation \cite{Hawking-1975cmp,Smerlak-S-2013prd},
 must either be removed from spacetime,
 meaning that $\manN$ is no longer topologically trivial,
 or alternatively that $\Jfree$ is not defined in all of $\manN$.

Next, 
 to break the gauge-free condition \eqref{CQ_pf_dH},
 we take the position that
 the only
 fundamental Maxwell's equations
 are \eqref{CQ_Max_F} and \eqref{CQ_Max_dJ},
 that is the closure of $\Fform$ and $\Jfree$.
Since equation \eqref{CQ_Max_HJ},
 and indeed $\Hform$ itself,
 would now not be considered fundamental,
 $\Hform$ may be considered as simply a potential for the current $\Jfree$.
As such,
 it will have its own gauge freedom,
 as discussed in the Introduction.
 Writing \eqref{eqn-vectorgauge} 
  in differential form notation,
 for any 1-form $\psi_{\subHP}\in\Gamma\Lambda^1 \manM$,
 where $\psi_{\subHP}$ encodes $\varphi_{\subHP}$ and $\emVec{A}_{\subHP}$,
 we have
\begin{align}
  \Hform \to \Hform + {\de} \psi_{\subHP}.
\label{CQ_H_Gauge}
\end{align}
This alternative interpretation of Maxwell's equations
 implies that similar to the usual 
 {1-form} potential $\Apotl$ for $\Fform$,
 the excitation field $\Hform$ is not measurable.
Since $\Hform$ is not defined absolutely,
 the Maxwell equation ${\de} \Hform=\Jfree$
 and the constitutive relation
 linking $\Hform$ to $\Fform$
 must be replaced by a constitutive relation
 relating the measurable quantities $\Fform$ and $\Jfree$. This might take the form of relating $\Jform$ to $\Fform$ and its derivatives, for example.
Thus we may interpret \eqref{CQ_Max_HJ} and \eqref{CQ_CR_vac} as
 two aspects of a \textit{single} constitutive relation for the Maxwell vacuum
\begin{align}
  {\de} {\hstar} \Fform
 -
  \Jfree
=
  0
.
\label{CQ_CR_J_D*F}
\end{align}
An alternative, 
 axion-like,
 constitutive relation
 might be given by
\begin{align}
  {\de} {\hstar} \Fform - \Jfree
=
  \psi\wedge\Fform
,
\label{CQ_CR_J_D*F_psiF}
\end{align}
where $\psi \in \Gamma \Lambda^1 \manM$ is a prescribed closed 1-form. 
For the electromagnetic potential $\Apotl$, 
 in \eqref{CQ_CR_J_D*F_psiF} we can
 write $\Hform={\hstar} \Fform+\psi\wedge \Apotl$
 but this does not define $\Hform$ uniquely. 
Likewise for a (non unique) $\phi\in\Gamma\Lambda^0\manM$
 with ${\de}\phi=\psi$
 one has $\Hform={\hstar} \Fform+\phi\, \Fform$.

When considering constitutive relations in a medium
 we distinguish the free current
 $\JFree \in \Gamma \Lambda^3 \manM$ from the bound current
 $\Jbdd \in \Gamma \Lambda^3 \manM$ representing the polarisation of
  the medium. The total current
 $\Jtot \in \Gamma \Lambda^3 \manM$ is given by $\Jtot=\JFree+\Jbdd$.
We set $\JFree=\Jfree$ in the above and describe the difference
 between the $\JFree$ and ${\de} {\hstar} \Fform$ as the bound current $\Jbdd$.
Thus we replace \eqref{CQ_Max_HJ}
 with
\begin{align}
  {\de} {\hstar} \Fform - \JFree
&=
  \Jbdd
,
\label{CQ_def_JB}
\end{align}
where $\Jbdd$ is related to $\Fform$ via another constitutive relation.
{For example in \eqref{CQ_CR_J_D*F}, $\Jbdd=0$, whereas in
\eqref{CQ_CR_J_D*F_psiF} $\Jbdd=\psi\wedge\Fform$.}
The currents $\JFree$ and $\Jbdd$ will be used in what follows to
encode the effects of charge non-conservation.


It is worth noting that these two apparently distinct cases
 allowing for non-conservation of charge are related
 by topological considerations --
 the choice of spacetime with a line or point removed,
 the {non-existence} of a well-defined $\Hform$,
 and the breaking of global charge conservation are all related
 to the deRham cohomology of the spacetime manifold\footnote{The
  $k$-th deRham cohomology $\Hform^k_\deRahm(\hat{\manM})$
  of the manifold $\hat{\manM}$ is defined to be the equivalence class
  of closed $k-$forms modulo the exact forms.
 In the topologically trivial case all the $\Hform^k_\deRahm(\hat{\manM})=0$,
  with $k>0$,
  and hence all closed forms are exact.
 In the language here,
  this implies that since $\Jfree$ is closed,
  ${\de}\Jfree=0$ there must exist
  a 2-form $\Hform\in\Gamma\Lambda^2\hat{\manM}$
  such that ${\de}\Hform=\Jfree$.
 In general $\Hform$ is not unique but it is globally defined.
 In the case of an evaporating black hole,
  the deRham cohomology $\Hform^3_\deRahm(\hat{\manM})=\Real$.
 Therefore even though $\Jfree$ is closed,
  it is not exact,
  i.e. there is no $\Hform \in \Gamma \Lambda^2 \hat{\manM}$
  such that ${\de}\Hform=\Jfree$,
  and thus $\Jfree$ need not be not globally conserved.
 A similar analysis is connected to/with magnetic monopoles.
 If we remove a world-line from spacetime,
  then the $\Hform^2_\deRahm(\hat{\manM})=\Real$.
 This implies that there need not exist an electromagnetic potential $\Apotl$,
  where ${\de}\Apotl = \Fform$.
 Hence $\int_{S^2} \Fform \ne 0$ where $S^2$ is a sphere at a
  moment in time enclosing the ``defect''.
}.

\section{Singularity} 
\label{ch_Singular}

In this section we construct an orientable manifold
 $\manM$ on which 
 \UPDATED{charge is \textit{not} globally conserved,
 even though (locally) $d\Jform=0$ everywhere on $\manM$}.
We start by assuming a flat spacetime with a Minkowski metric,
 except with the significant modification that
 a single event
 $\eventZero$ has been removed;
 i.e. $\manM=\Real^4\backslash \eventZero$.
This spacetime $\manM$ is sufficient to
 demonstrate our mathematical and physical arguments for
 charge conservation failure --
 but without introducing any of the additional
 complications of (e.g.) the Schwarzschild black-hole metric.
As already noted in our Introduction, 
 the discussion here is entirely separate
 from and prior to any assumptions about cosmic censorship,
 or any imagined model of the singularity behaviour.

\subsection{Charge conservation}
\label{ch_SCQ}

\begin{figure}
\centering
\resizebox{0.90\columnwidth}{!}{
\includegraphics{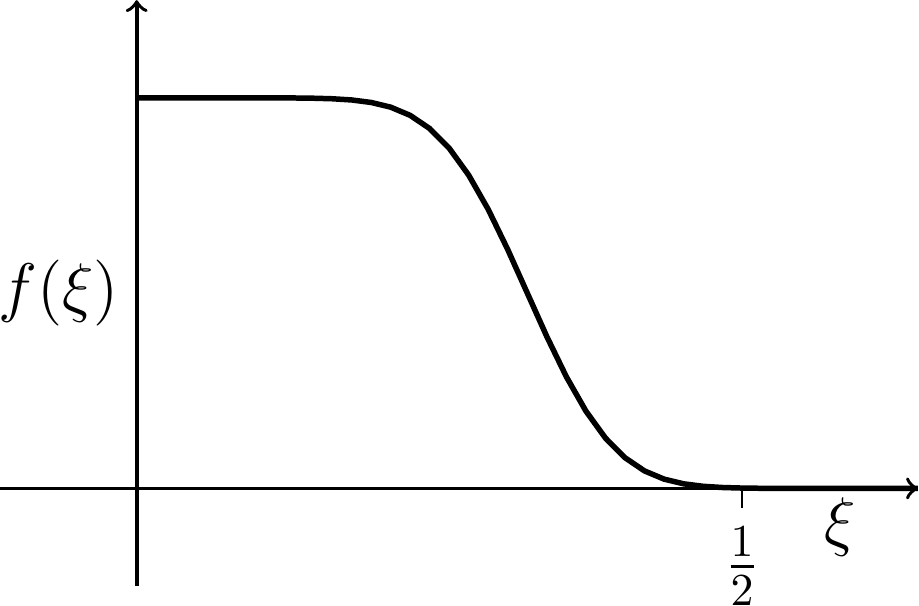}
}
\caption{A bump function ${f(\xi)}$ which we used to construct a smooth
  current density. The function is 
  completely flat for $\xi>\tfrac12$ and for $\xi$ near zero.}
\label{fig_Bump}
\end{figure}

Let $(t,x,y,z)$ be the usual Cartesian coordinate system
 with $\eventZero=(0,0,0,0)$ and let $(t,r,\theta,\phi)$ be
 the corresponding spherical coordinates\footnote{Let $\manM$ have
   signature $(-,+,+,+)$ and orientation ${\hstar} 1={dt}\wedge {dx}\wedge {dy}\wedge {dz}$.}. 
Set $\Real^{+}=\Set{r\in\Real|r\ge0}$.
Let us construct the smooth 3-form current density $\Jfree$
 defined throughout $\manM$ as
\begin{equation}
  \begin{aligned}
    \Jfree
  =
    \begin{cases}
      0 & \text{ for } t\le 0,
\\
      \Jfree^{+} &\text{ for }t > 0.
    \end{cases}
  \end{aligned}
\label{eg_J}
\end{equation}
This $\Jfree^{+}$ is then defined using
 a 
 function ${f}:\Real^{+}\to\Real^{+}$
 ${{f}(\xi)} \ge 0$ 
 \UPDATED{for $0 \le \xi < \tfrac12$,}
 ${{f}(\xi)}=0$
 for $\xi \ge \tfrac12$ and all the derivatives ${{f}^{(n)}(0)}=0$ for
 $n\ge1$. 
Such functions are usually called bump functions,
 an example of which is shown in fig. \ref{fig_Bump}. 
\UPDATED{Here $\xi$
 is simply the argument of the function $f$
 (and also of the function $h$ below), 
 and it is replaced by $r/t$ when the function is used to define fields.}
We then have
\begin{align}
  \Jfree^{+}
&=
  \frac{1}{t^3}\, 
  {{f}\!\left(\frac{r}{t}\right)}
  \, 
  {dx}\wedge {dy}\wedge {dz}
\nonumber
\\
&\quad
 -
  \frac{1}{t^4}
  {{f}\!\left(\frac{r}{t}\right)}
  {dt} \wedge
  \left( x\,{dy}\wedge {dz} + y\,{dz}\wedge {dx} + z\,{dx}\wedge {dy} \right)
.
\label{eg_J+}
\end{align}
The first term on the right hand side of \eqref{eg_J+}
 represents the charge density,
 while the second term represents the current density\footnote{\UPDATED{One
   may think of our proposed $\Jfree$
   as an application of deRham's second theorem. 
  Since the unit 3-sphere about the origin is a 3-cycle
   which is \emph{not} a boundary,
   deRham's second theorem states that for any real valued ${Q}$
   there always exists a 3-form $\mForm{\omega}$
   on $\manM$ such that $\int_{S^3} \mForm{\omega} = {Q}$. 
  Here our $\mForm{\omega}$ is $\Jfree$,
   which is chosen so that after the initial ``impulse'' at $\eventZero$, 
   it subsequently respects causality.}}.
We note that as expressed in the Cartesian coordinates of \eqref{eg_J+},
 $\Jform^{+}$ is well defined at the spatial origin for $t>0$. 
In spherical polar coordinates we have
~
\begin{align}
  \Jfree^{+}
=
  \sin(\theta)
  ~
  {{f}\!\left(\frac{r}{t}\right)}
  \left(
    \frac{r^2}{t^3}\, 
    {dr}
   - 
    \frac{r^3}{t^4} \, {dt} 
  \right)
\wedge {d\theta}\wedge {d\phi}
.
\label{eg_J+_alt}
\end{align}

To establish charge conservation on all of $\manM$ with ${\de}\Jfree=0$,
we first note, {from \eqref{eg_J+_alt}}, that ${\de} \Jfree^{+}=0$ for $r
> 0$ and $t>0$.  Since $f^{(1)}(0)=0$ then, {from \eqref{eg_J+}},
${\de} \Jfree^{+}=0$ for $t>0$ and $r= 0$.  Moreover, since $\Jfree=0$ for
$t< 0$, we have that ${\de} \Jfree=0$ for $t<0$.  For the hypersurface
$t=0$, we note that about any point for which $r\ne 0$ there exists an
open set in $\manM$ on which $\Jfree = 0$ and hence
$\left.\Jfree\right|_{t=0}=0$.  Thus ${\de}\Jfree=0$ on all of $\manM$.

Physically,
 \eqref{eg_J} and \eqref{eg_J+_alt} seems to represent
 a $\delta$-function of charge ${\Charge}$
 appearing at the origin at $t=0$, 
 and then spreading out spatially from $\eventZero$ into $\manM$; 
 where
~
\begin{align}
  {\Charge}
&=
  4\pi \int_0^\infty \xi^2 {{f}(\xi)} {d\xi}
.
\label{eqn-Qchargeintegral}
\end{align}
However, 
 the spacetime origin is {\textit{not}} an event in $\manM$,
 and the ${\Charge}$'s appearance at $t=0$ does not induce ${\de}J\ne 0$
 at some event in $\manM$. 
We see that
 the total charge is zero for the constant-time hypersurfaces\footnote{In fact any Cauchy hypersurface with $t<0$ suffices here.} with $t<0$,
 but for the constant-time hypersurfaces
 with $t>0$ the charge is ${\Charge}$.

Similarly,
 over a region such as that shown in 
 figs. \ref{fig_consCharge} and \ref{fig_singular}
 we have that $\int_\manU \Jform \ne 0$. 
Charge is therefore not conserved in $\manM$,
 despite the fact that ${\de}\Jform = 0$ everywhere in $\manM$.

\begin{figure}
\centering
\resizebox{0.90\columnwidth}{!}{
\includegraphics{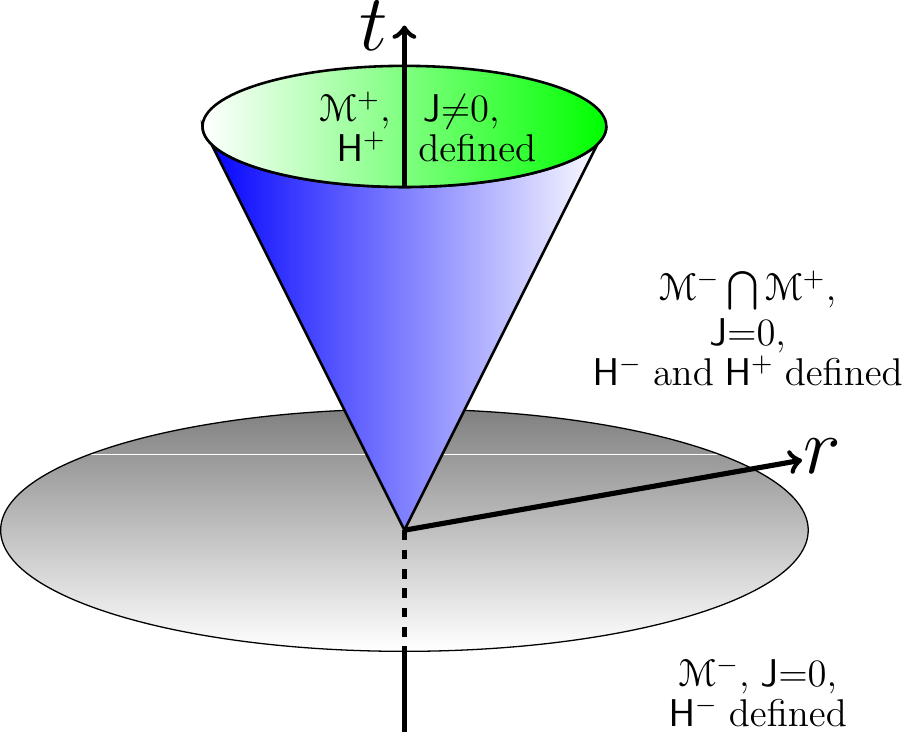}
}
\caption{A spacetime $\manM$ in which charge is not conserved. 
The forward cone,
 lying within the lightcone of the excised event at $\eventZero$
 has non-zero charge, 
 whereas the remainder of the spacetime is uncharged.  
We show the regions $\manM^{+}$ and $\manM^{-}$
  where the excitation 2-forms $\Hform^{+}$ and $\Hform^{-}$ are defined.
}
\label{fig_Mpm}
\end{figure}

Now since $\manM$ is topologically non-trivial, 
 it is impossible to find a single $\Hform$ such that ${\de} \Hform =0$.
This \textit{must} be the case since if it were not,
 then we could apply \eqref{CQ_pf_dH}
 to establish ${\Charge}=0$.
Nevertheless we \textit{can} find two fields $\Hform^{+}$ and $\Hform^{-}$
 with intersecting domains $\manM^{+}$ and $\manM^{-}$
 such that on the intersection,
 $\Hform^{+}= \Hform^{-} + {\de}\psi$,
 i.e they differ by just a gauge,
 as per \eqref{CQ_H_Gauge}.
Let
\begin{align}
  \manM^{+} = \manM\backslash\Set{t<0}
  ~~\textrm{and}~~
  \manM^{-} = \manM\backslash\Set{t>0 ~\textrm{and}~ r<\tfrac12 t}
,
\label{eg_def_Mpm}
\end{align}
and with $\Hform^{+}\in\Gamma\Lambda^2 \manM^{+}$ 
 and $\Hform^{-} \in \Gamma\Lambda^2 \manM^{-}$,
 let
\begin{equation}
\begin{aligned}
  \Hform^{+} 
&=
  {{h}\left(\frac{r}{t}\right)}
  \left(x\,{dy}\wedge {dz} + y\,{dz}\wedge {dx} + z\,{dx}\wedge {dy} \right)
\nonumber
\\
&=
  {{h}\left(\frac{r}{t}\right)}
  \left(r^3\sin\theta\, {d\theta} \wedge {d\phi} \right)
,
\\
  \Hform^{-} &= 0,
\end{aligned}
\label{eg_def_Hpm}
\end{equation}
as depicted in fig. \ref{fig_Mpm}. 
Here
\begin{align}
  {{h}(\xi)}
&=
  \frac{1}{\xi^3} \int_0^\xi {{f}({\hat \xi})}\, \hat{\xi}^2 d\hat{\xi}
.
\label{eg_def_h}
\end{align}
Since ${{h}(\xi)}$ is smooth about $\xi=0$ and
\begin{align}
  {{h}(\xi)}
=
  \frac{{\Charge}}{4\pi \xi^3}
 \quadtext{for}
   \xi>\tfrac12
,
\label{eg_h_r>1}
\end{align}
then on the intersection 
 $\manM^{+}\inter \manM^{-}$$
  =
   \Set{ \left.\left(t,r,\theta,\phi\right)\,\right|\,r>t>0}$ we have
\begin{align}
  \Hform^{+} 
=
  \frac{{\Charge}}{4\pi} \sin(\theta){d\theta}\wedge {d\phi}
\qquadand
  \Hform^{-} 
=
  0
.
\label{eg_Hpm_inter}
\end{align}
Therefore 
 $\Hform^{+} 
= \Hform^{-}-({\Charge}/4\pi){d\left(\cos(\theta)\,{d\phi}\right)}$
 for the region 
 $\Set{\left.(t,r,\theta,\phi)\,\right|\,r>t>0,\theta\ne0,\pi}$. 
Choosing other patches of the $(\theta,\phi)$
 sphere we can find other gauge fields $\psi$
 such that $\Hform^{+} = \Hform^{-} + {\de}\psi$. 
In this example,
 there is no global $\Hform$-field,
 and both $\Hform^{+}$ and $\Hform^{-}$ fail in distinct regions of $\manM$. 
This strongly suggests that $\Hform$
 need not have absolute physical significance,
 unlike $\Fform$.

The new gauge freedom for $\Hform$
 suggests the possibility of further generalisations 
 to the vacuum constitutive relations.
 These could now go beyond rather prescriptive vaccum models 
 such as e.g. the Euler-Heisenberg or Bopp-Podolsky ones
 in \eqref{CQ_CR_EH} and \eqref{CQ_CR_BP}, 
 whose Lagrangian formulations insist on a unique $\Hform$.

\section{Polarisation of the vacuum}
\label{ch_Pol}

We now stay with the same scenario as in the previous section, 
 but instead apply
 the bound current version of Maxwell theory
 as given by \eqref{CQ_def_JB},
 interpreting $\Jbdd$ as representing the polarisation of the vacuum.  
It is known from quantum field theory that
 vacuum polarization occurs naturally for intense fields,
 with the first order correction
 to the excitation 2-form given by \eqref{CQ_CR_EH}.  
Indeed,
 the strong magnetic fields associated with magnetars
 are known to induce non-trivial
 dielectric properties on vacuum \cite{Lai-H-2003aj}. 
An alternative model for the polarization of the vacuum
 is given by the Bopp-Podolski theory of electromagnetism, 
 as outlined in \eqref{CQ_CR_BP}. 
However in these cases the bound currents
 $\Jbdd^{\!\!\!\!\textup{EH}}=d\Hform_{\textup{EH}}-{\de}{\hstar}\Fform$ and 
 $\Jbdd^{\!\!\!\!\textup{BP}}=d\Hform_{\textup{BP}}-{\de}{\hstar}\Fform$ 
 correspond to a well defined
 excitation 2-form $\Hform$ and therefore must conserve charge, 
 regardless of topology. 
Nevertheless we are still free to consider
 more general versions of $\Jbdd$ which are not exact and 
 contain more than just those corrections.

Since ${\hstar} \Fform$ is well defined,
 and ${\de}{\hstar} \Fform=\JFree+\Jbdd$,
 one can use the argument \eqref{CQ_pf_dH},
  replacing $\Hform$ with ${\hstar} \Fform$,
 to conclude that $\JFree+\Jbdd$ is globally conserved.
We now examine whether it is
 necessary for $\JFree$ and $\Jbdd$ to be 
 \UPDATED{globally conserved \textit{independently}}.
If $\Hform$ is well defined then from \eqref{CQ_Max_HJ}
 (which now becomes $d\Hform=\JFree$)
 ${\de}\JFree=0$ and hence ${\de}\Jbdd=0$.
Under these circumstances,
 we find that $\JFree$ by itself is globally conserved,
 and likewise for $\Jbdd$ by itself.

\begin{figure}
\centering
\resizebox{0.90\columnwidth}{!}{
\includegraphics{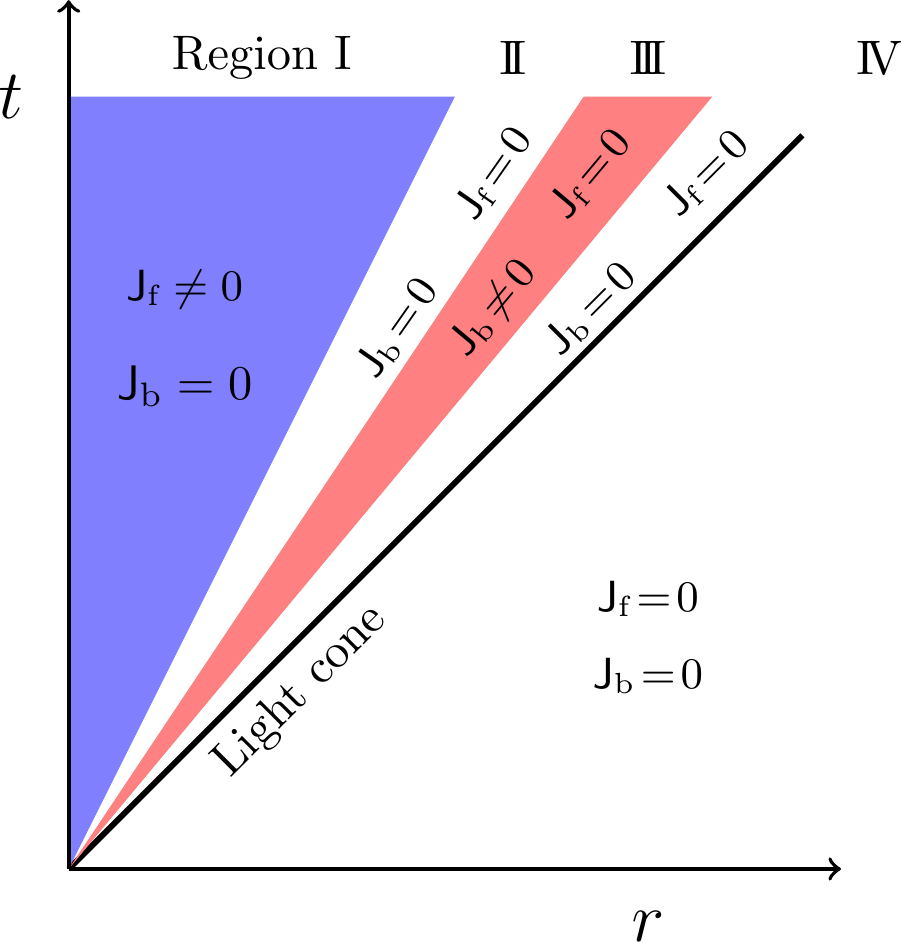}
}
\caption{Domains in a spacetime $\manM$
 where $\JFree$ (blue) and $\Jbdd$ (red) may be non-zero;
 note in particular that the supports of $\JFree$ and $\Jbdd$ do not intersect.
For completeness,
 we show several different regions
 where the various possible combinations of
 zero and non-zero $\JFree$ and $\Jbdd$ hold,
 although alternative (and simpler)
 scenarios are possible.
Note that 
 Region $\RegI$ matches the cone shown on fig. \ref{fig_Mpm},
 and
Region $\RegIV$ encompasses both a section
 above the light line, 
 as well as below.
}
\label{fig_J_Jbdd}
\end{figure}

In the following,
 we demand only that $\JFree$ and hence $\Jbdd$ are closed:
 $d\JFree = d\Jbdd = 0$,
 and do not insist that they are globally conserved independently.
This requires us to abandon the concept
 of a global macroscopic well-defined $\Hform$,
 and to express the constitutive relations for our spacetime
 using the microscopic bound current,
 $\Jbdd$.
Unlike $\Fform$, 
 the excitation $\Hform$ cannot be measured directly using either
 the Lorentz force equation or the Aharonov-Bohm effect.

To demonstrate our replacement of $\Hform$, 
 we now show that we can replace it with a bound current. 
\UPDATED{We start by choosing an appropriate $\Fform$,
 setting} 
\begin{align}
  \Fform 
&=  
  h\!\left(\frac{r}{t}\right)
  \chi\!\left(\frac{r}{t}\right)
  \heavi(t) 
  \, 
  r \, {dt} \wedge {dr}
=
  {{R}\left(\frac{r}{t}\right)}
  \,
  r \, {dt} \wedge {dr}
\label{eg_def_F}
,
\end{align}
where 

\begin{description}

\item[(i)] \UPDATED{$h(\xi)$
 is defined from \eqref{eg_def_h},
 i.e. $h(\xi) 
  = \frac{1}{\xi^3}\int_0^\xi f({\hat{\xi}}){\hat \xi}^2d{\hat \xi}$.}

\item[(ii)] \UPDATED{$f: {\mathbb R}^+\rightarrow {\mathbb R}^+$
 is again a bump function satisfying $f(\xi) \ge 0$  for $0 < \xi < 1/2$,
 $f(\xi) =   0$ for $\xi \ge 1/2$,
 and $f^{(n)}=0$ for $n \ge 1$.}

\item[(iii)] \UPDATED{$\heavi:\Real\to\Real$ 
 is the Heaviside step function.}

\item[(iv)] \UPDATED{${\chi}:\Real^{+}\to\Real$ 
 is a bump function
 with ${{\chi}(\xi)}=1$ for $0\le \chi<\tfrac23$
 and ${{\chi}(\xi)}=0$ for $\xi>\tfrac56$.}

\end{description}


Clearly ${\de}\Fform=0$. 
The scalar factor ${{R}(\xi)}$ on the right hand side of
\eqref{eg_def_F} has the following properties
\begin{eqnarray}\label{triplefunction}
  {{R}\left(\frac{r}{t}\right)}
&=\!&
  \left\{
    \begin{array}{ll@{\quad}l}
    \eqstrut
    ~~ 0 
    & \text{ for } t<0,
    & 
\\
    \eqstrut 
    ~~ {{h}\!\left(\frac{r}{t}\right)} 
    & \text{ for }0\le r<\tfrac12 t, 
    & \textrm{Region~{\RegI}} 
\\
    \eqstrut
    \frac{\Charge}{4\pi r^3} 
    & \text{ for }0<\tfrac12 t < r<\tfrac23 t,\!
    & \textrm{Region~{\RegII}} 
\\
    \eqstrut
    {{\chi}\!\left(\frac{r}{t}\right)}
    \frac{\Charge}{4\pi r^3} \!
    & \text{ for }0<\tfrac23 t < r<\tfrac56 t,\!
    & \textrm{Region~{\RegIII}} 
\\
    \eqstrut
    ~~ 0
    & \text{ for }0<\tfrac56t<r, 
    & \textrm{Region~{\RegIV}} 
  \end{array}
  \right.
\nonumber
\\
  \qquad
\end{eqnarray}
where
 $\Charge$ is given by \eqref{eqn-Qchargeintegral}, 
 and
 the regions {\RegI} to {\RegIV} are shown in fig. \ref{fig_J_Jbdd}. 
These regions contain a selection
 of the possible combinations of $\JFree$ and $\Jbdd$.
We can then set the constitutive relation
 to be that of \eqref{CQ_def_JB},
 with $\JFree, \Jbdd\in\Gamma\Lambda^3 \manM$ 
 being independently conserved, 
 but only in a \textit{local} sense, 
 not globally.
For our example, 
 they are 
 respectively given by
\begin{equation}
\begin{aligned}
  \JFree
=
  \left\{
  \begin{array}{ll@{\quad}l}
  \displaystyle
  d{\hstar} \Fform 
  & \text{ for }0<r< \tfrac12t,
    \hfill \qquad\qquad\qquad \text{in Region~{\RegI}}
\\
  \eqstrut
  0 
  & \text{ for } t<0 \text{ and } \tfrac12t<r< t,
\\
  & \hfill \text{in Regions~{\RegII}, {\RegIII} and {\RegIV}}
  \end{array}
  \right.
\end{aligned}
\label{eg_def_Jbb}
\end{equation}
and
~
\begin{equation}
\begin{aligned}
\Jbdd
=
 \left\{
  \begin{array}{ll@{\quad}l}
   0 
   & \text{ for } t<0 \text{ and } 0<r< \tfrac23t\text{ and } 0<\tfrac56t<r,~
\\
   & \hfill \text{in Regions~{\RegI}, {\RegII} and {\RegIV}}
\\
   \eqstrut
   \displaystyle
   d{\hstar} \Fform \!
   & \text{ for }
   0<\tfrac23 t < r<\tfrac56 t, 
   \hfill \text{in Region~{\RegIII}}
  \end{array}
 \right.
\end{aligned}
\label{eg_def_Jbdd}
\end{equation}
or explicitly as
\begin{equation}
\begin{aligned}
  \JFree
=
  \left\{
  \begin{array}{ll@{\quad}l}
  \displaystyle
    {{h}'\left(\frac{r}{t}\right)}
  \left[\left(\frac{r^3}{t}+3r^2\right)\,{dr} - \frac{r^4}{t^2} {dt}\right]
  \wedge\sin(\theta) {d\theta}\wedge {d\phi}
\\
  \hfill \text{ for } 0<r< \tfrac12t,
  \qquad \text{in Region~{\RegI}}
\\
  \eqstrut
  \displaystyle
  0
  \hfill \text{ for }
  \tfrac12t  < r< t,
  \qquad \text{in Regions~{\RegII}, {\RegIII} and {\RegIV}}
  \end{array}
  \right.
\end{aligned}
\label{eg_res_Jbb}
\end{equation}
and
~
\begin{equation}
\begin{aligned}
  \Jbdd
=
  \left\{
  \begin{array}{ll}
  0 
  \hfill \qquad\text{ for } t<0 \text{ and } 0<r< \tfrac23t
    \text{ and } 0<\tfrac56t<r,
\\
  \hfill\qquad\text{in Regions~{\RegI}, {\RegII} and {\RegIV}}
\\
  \eqstrut
  \displaystyle
  \frac{{\Charge}}{4\pi} {{\chi}'\left(\frac{r}{t}\right)}
  \left[\frac{1}{t}\,{dr} - \frac{r}{t^2} {dt}\right]
    \wedge\sin(\theta) {d\theta}\wedge {d\phi}
\\
  \hfill\qquad\text{ for }
    0<\tfrac23 t < r<\tfrac56 t,
  \qquad\text{in Region~{\RegIII}}
.
  \end{array}
  \right.
\end{aligned}
\label{eg_res_Jbdd}
\end{equation}

The occurrence of a bound charge of single sign
 over an extended region of space may seem rather unusual.
However,
 this can be realised in a dielectric
 with a continuously varying permittivity.
For example,
 a constant bound charge density $\lambda$ can be obtained with
 the dielectric varying as
\begin{align}
  {{\epsilon}(z)}
&=
  \epsilon_0
 +
  \epsilon_0
  \frac{P_0 - \lambda\,z}
       {\epsilon_0\,V_0+\lambda\,z}
,
\end{align}
which gives
~
\begin{align}
  {{E}(z)}
=
  \frac{\lambda\,z}{\epsilon_0}+V_0
\quadand
  {{P}(z)}
=
  P_0
 -
  \lambda\,z
,
\end{align}
for constants $V_0$ and $P_0$.

In writing \eqref{eg_def_F} 
 the distinction between free and bound current densities,
  as arise in the subsequent calculations,
 is introduced artificially.  
So whilst our introduced example is certainly artificial,
 it can be taken to be representative and illustrative
 of a scenario in which the
 2-form field $\Hform$ is no longer globally defined;
 but that the
 sum of free and bound charge densities in vacuum
 is globally conserved,
 while the two types of charge are not collectively globally conserved,
 and exist independently in disjoint regions of space.
This echoes the previous section, 
 but here we use the non-exactness of $\Jform$
 rather than the gauge freedom for $\Hform$; 
 thus suggesting generalisations 
 to the charge and polarization properties
 of the vacuum constitutive relations.

\section{Wormhole} 
\label{ch_Worm}

\begin{figure}[!ht]
{
\resizebox{0.90\columnwidth}{!}{\includegraphics{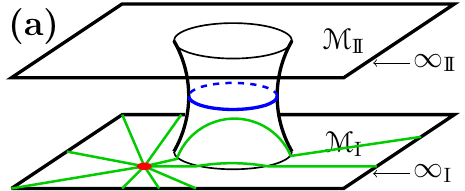}}
~\\
~\\
\resizebox{0.90\columnwidth}{!}{\includegraphics{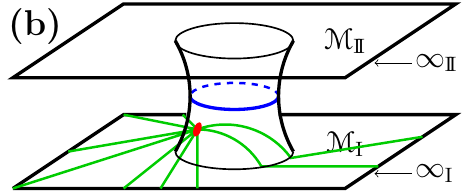}}
~\\
~\\
\resizebox{0.90\columnwidth}{!}{\includegraphics{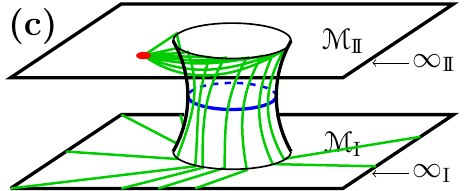}}
}
\caption{
A depiction of the standard interpretation
 of field line behaviour 
 as a charge {$q$}
 moves from the lower universe $\manM_\RegI$ (a), 
 through the wormhole (b),
 and into the upper universe $\manM_\RegII$ (c) --
 all the field lines (green)
 from the point charge (red dot)
 must remain attached to $\infty_\RegI$.
Note that despite the similarity in relative charge-wormhole positions
 between the starting point in (a) 
 and the end point in (c), 
 the field line configuration is very different.
Consequently,
 as the charge moves ever further
 from the wormhole mouth in $\manM_\RegII$,
 the effective charges in the biverse would number three and not one:
 $\manM_\RegI$ has a wormhole mouth with 
 field lines exiting it on the way to infinity, 
 and which integrate to $+{q}$, 
 whilst $\manM_\RegII$ has an overall dipole-like field between
 an effective charge $-{q}$ on the wormhole mouth
 and the free charge $+{q}$.
}
\label{fig_Wormhole_StandardTail}
\end{figure}

A wormhole \cite{Morris-T-1988ajp}
 is another example of a non-trivial spacetime,
 although in this case it is the first deRham cohomology
 which does not vanish, 
 $\Hform^1_\deRahm=\Real$.
In this scenario we do \textit{not} break conservation of global charge, 
 but instead address the issue of whether a wormhole
 necessarily 
 gains the charge of any matter passing through it.
One simple way of describing this standard viewpoint is to note that 
 the usual process of drawing field lines for a charge, 
 as it moves, 
 forbids them from 
 swapping their end-points from one place to another.
This means that a positive charge
 passing through a wormhole ``drags'' its 
 field lines behind it like a tail, 
 and the resulting collection of field lines re-entering the wormhole
 looks like a negative charge, 
 and 
 then as they exit the other side they look like a positive charge; 
 as depicted in fig. \ref{fig_Wormhole_StandardTail}.

\begin{figure}[!ht]
\resizebox{0.90\columnwidth}{!}{
\includegraphics{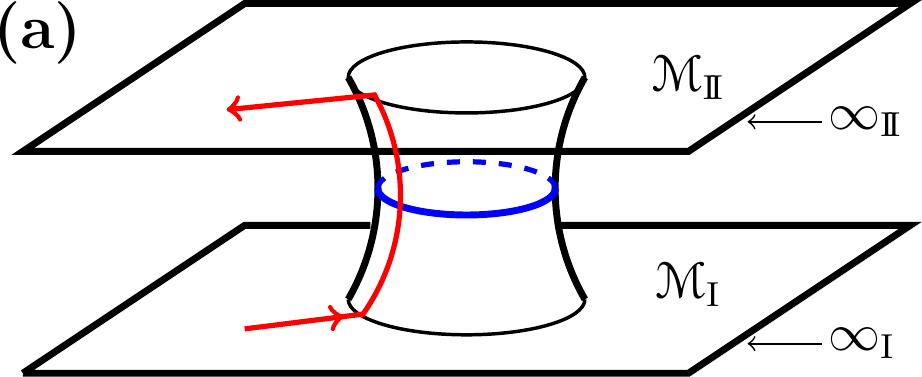}
}\\
~\\
\resizebox{0.330\columnwidth}{!}{
\includegraphics{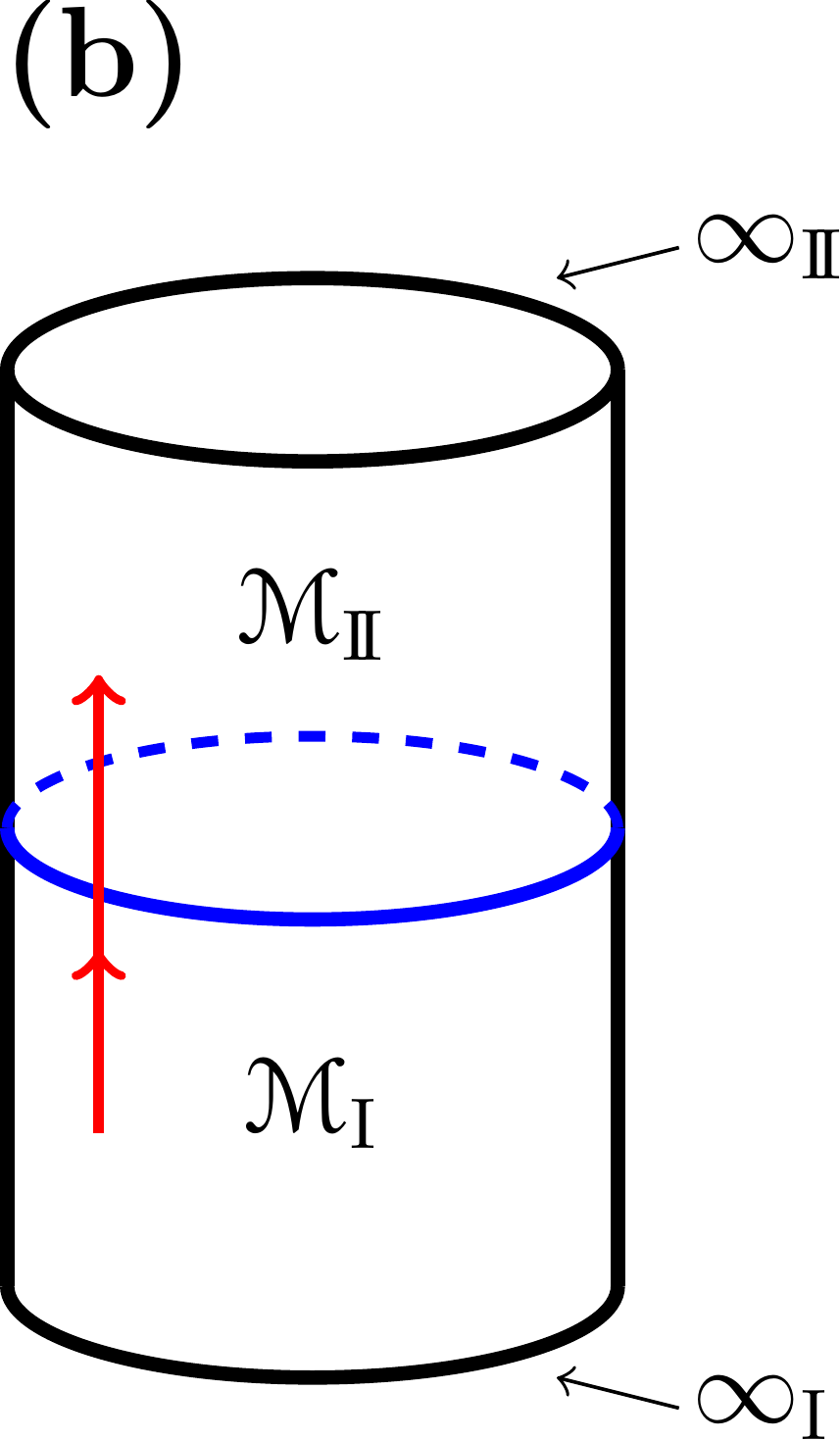}
}
\qquad
\resizebox{0.470\columnwidth}{!}{
\includegraphics{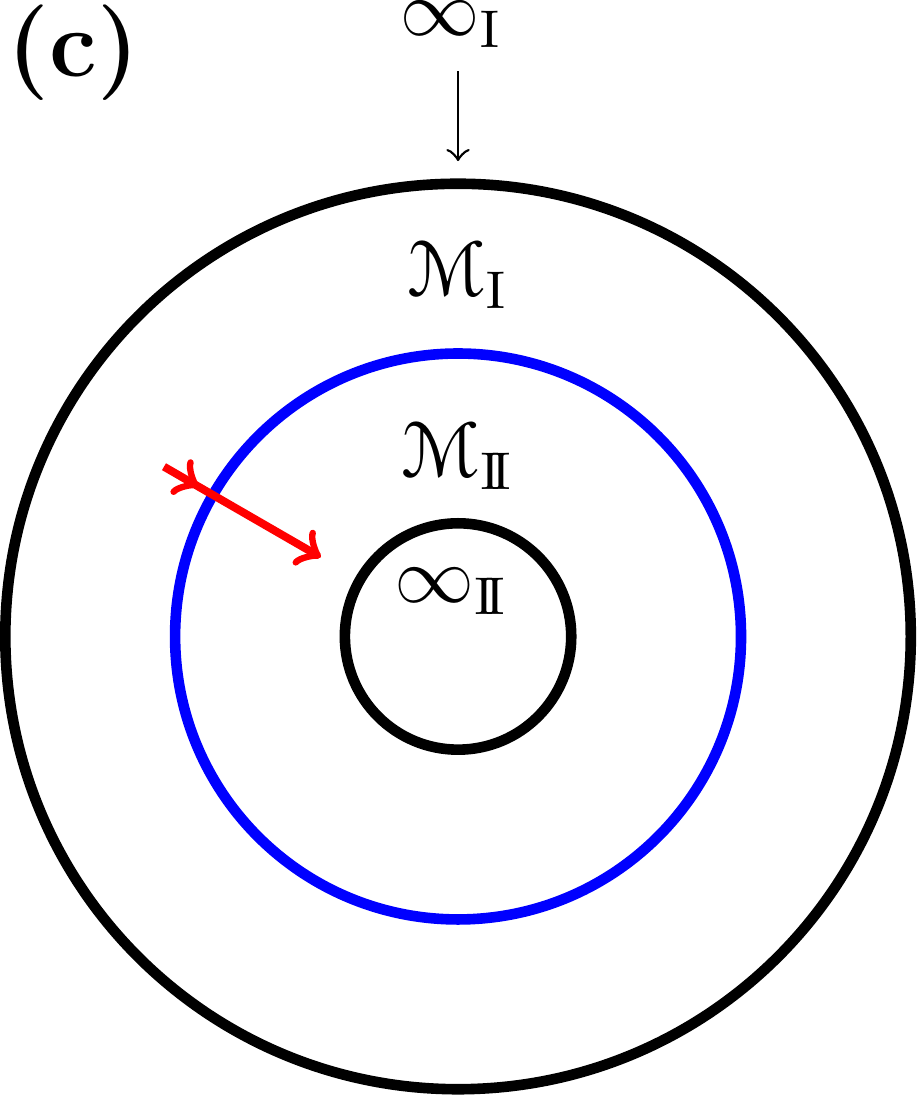}
}
\caption{Deforming a ``biverse'', 
 i.e. two {otherwise distinct} universes interconnected by a wormhole,
 into an annulus. 
When deforming from (a) into (b), 
 the wormhole's throat (blue line) is unchanged, 
 whilst the top ($\manM_\RegII$) and bottom
  ($\manM_\RegI$) universes are deformed into cylinders. 
The final stage from (b) to (c)
 requires opening out the cylinders into two nested annuli, 
 which form a single annulus with the throat
 demarking the join.
The red line in each diagram represents the path for a
  point charge leaving $\manM_\RegI$ and entering $\manM_\RegII$.}
\label{fig_Wormhole}
\end{figure}

The proof of conservation of charge is similar to the arguments of
 \eqref{CQ_pf_dJ} and \eqref{CQ_pf_dH} but in this case the two
 arguments have different interpretations.  
First we note from fig. \ref{fig_Wormhole}
 that spatially the wormhole is topologically equivalent
 to a 3-dimensional annulus,
 i.e. a 3-ball with an inner 3-ball removed. 
The inner and outer 2-spheres $\infty_\RegI$
 and $\infty_\RegII$ represent spacelike infinity in the two
 universes $\manM_\RegI$ and $\manM_\RegII$. 
Between the two 2-spheres there is a concentric sphere which is the throat. 
Although geometrically the throat is the minimum size sphere
 which connects the two universes, 
 topologically there is nothing special about the throat,
 and here we take it as a convenient place to talk about
 where one passes from one universe to the other.
{Further,
 since each universe ($\manM_\RegI$ or $\manM_\RegII$)
 has its own infinity ($\infty_\RegI$ or $\infty_\RegII$),
 there are two ways to be arbitrarily far from the wormhole, 
 and thus there are two possible destinations for the field lines
 of a charge.}

\begin{figure}[!ht]
\resizebox{0.90\columnwidth}{!}{\includegraphics{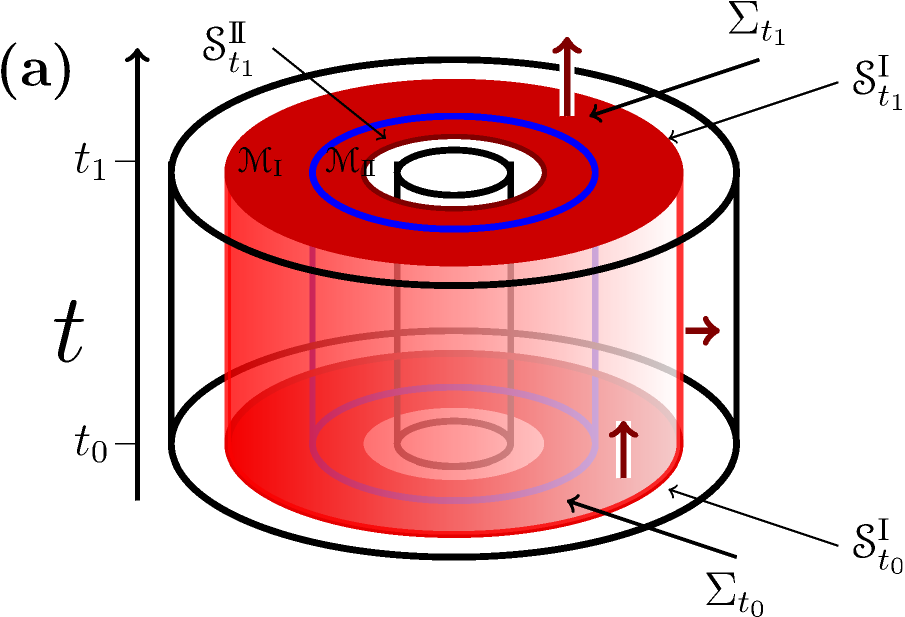}
}\\
~\\
\resizebox{0.90\columnwidth}{!}{\includegraphics{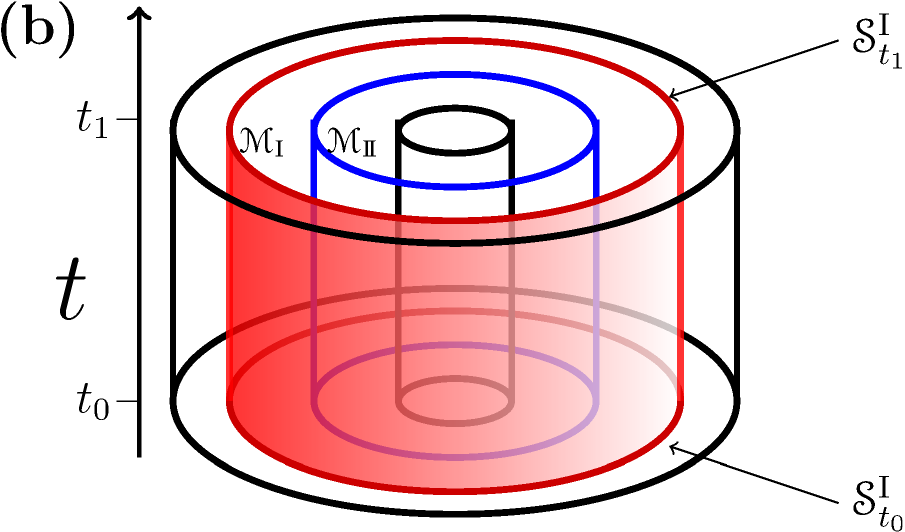}
}
\caption{Integration submanifolds for our wormhole example
 of fig. \ref{fig_Wormhole}.
The upper diagram (a) is for integrating $\Jform$,
 whereas the lower diagram (b) is for integrating $\Hform$. 
The orientation is shown in dark red.}
\label{fig_Wormhole_int_regions}
\end{figure}

Consider the 3-dimensional region $\manSigma$ bounded
 by the spheres $\manS^\RegI$ and $\manS^\RegII$. 
Then,
 as illustrated in fig. \ref{fig_Wormhole_int_regions}(a), 
 $[t_0,t_1]\times\manSigma$ is a 4-dimensional region
 bounded by $\manSigma_{t_0}$, $\manSigma_{t_1}$,
 $[t_0,t_1]\times \manS^\RegII$
 and $[t_0,t_1]\times \manS^\RegI$.
Then
\begin{align}
  0
&=~~
  \int_{[t_0,t_1]\times\manSigma} {\de}\,\Jfree
=~~
  \int_{{\boundary}\left(\rule{0.0em}{0.7em}[t_0,t_1]\times\manSigma\right)}
    \Jfree
\nonumber
\\
&=
  \int_{\manSigma_{t_1}} \Jfree
 -
  \int_{\manSigma_{t_0}} \Jfree
 +
  \int_{[t_0,t_1]\times \manS^\RegII} \Jfree
 +
  \int_{[t_0,t_1]\times \manS^\RegI} \Jfree
.
\label{Worm_J_balance}
\end{align}
This states that the total charge in $\manSigma$ at time $t_1$
 equals the charge in $\manSigma$ at time $t_0$
 plus any charge that enters (or leaves)
 via $\manS^\RegII$ and $\manS^\RegI$. 
We cannot let $\manS^\RegI$ go to infinity
 as then it would disappear from 
 the right hand side of \eqref{Worm_J_balance}
 and Stokes' theorem will no longer apply.

\begin{figure}[t]
\resizebox{0.90\columnwidth}{!}{
\includegraphics{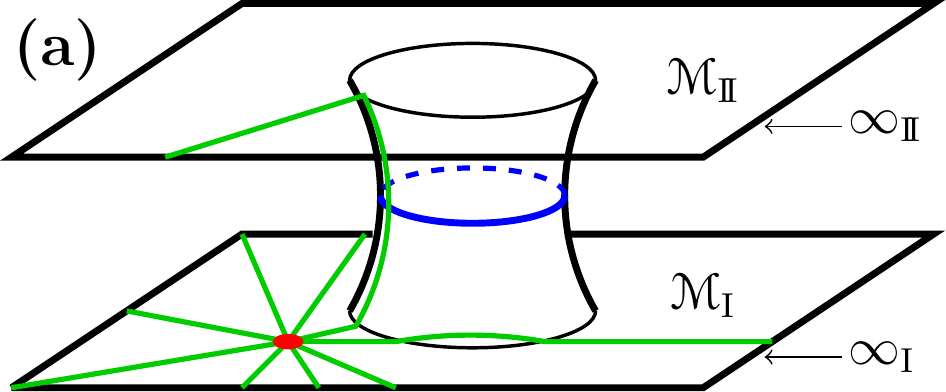}
}
\\
~\\
\resizebox{0.90\columnwidth}{!}{
\includegraphics{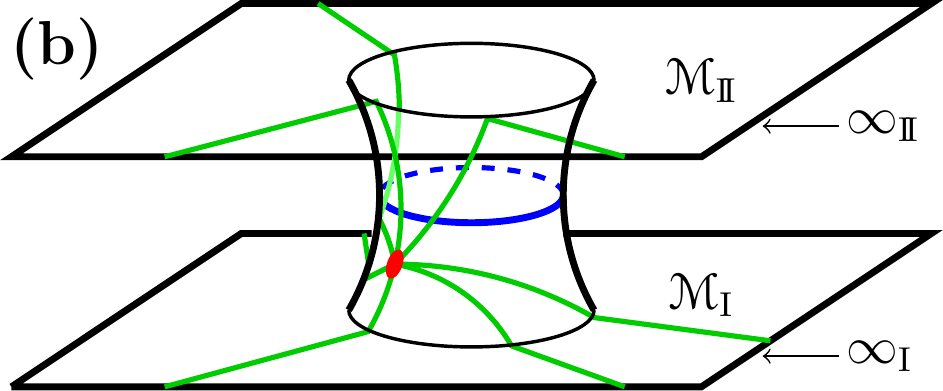}
}
\\
~\\
\resizebox{0.90\columnwidth}{!}{
\includegraphics{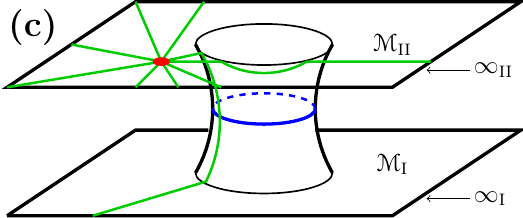}
}
\caption{Depiction of the alternative field line behavior
when allowing for the topology and the new freedom for $\Hform$.
Field lines (green) in the biverse 
 all start a the point charge (red dot),
 but now have a choice of infinity at which to terminate.
Initially, 
 when the charge is (a) well within universe $\textup{I}$, 
 relatively few field lines reach through the wormhole into 
 universe $\RegII$ and hence off to $\infty_\RegII$, 
 but as the 
  point charge moves into to the throat (b)
  more of the field lines will reach $\infty_\RegII$.
In (c)
 we see that after the charge has moved fully into universe $\textup{II}$, 
 in a position mirroring that in (a), 
 the arrangement of field lines is also mirrored, 
 in distinct contrast with the standard treatment of
 fig. \ref{fig_Wormhole_StandardTail}.
}
\label{fig_Wormhole_fieldlines}
\end{figure}

Another issue arises when considering the second proof of charge conservation
 (cf. \eqref{CQ_pf_dH}). 
If $\Hform$ is a well-defined $2$-form field we may integrate it
 over the 3-dimensional timelike hypersurface 
 $[t_0,t_1]\times \manS^\RegI$, 
 fig. \ref{fig_Wormhole_int_regions}(b). 
Setting
 ${\Charge}^\RegI_t=\int_{\manS^\RegI_t} \Hform$ as the charge inside
 $\manS^\RegI$ at time $t$, we have
\begin{align}
  {\Charge}^\RegI_{t_1}
 -
  {\Charge}^\RegI_{t_0}
&=~~
  \int_{\manS^\RegI_{t_1}} \Hform -
  \int_{\manS^\RegI_{t_0}} \Hform
=~~
 \int_{{\boundary}\left([t_0,t_1]\times \manS^\RegI\rule{0.0em}{0.7em}\right)}
 \Hform
\nonumber
\\
&=~~
  \int_{[t_0,t_1]\times \manS^\RegI} {\de} \Hform
=~~
  \int_{[t_0,t_1]\times \manS^\RegI} \Jform
.
\label{Worm_H_balance}
\end{align}
Thus if $\Hform$ is well-defined,
 and no current passes through $\manS^\RegI$,
 then ${\Charge}_t^\RegI$ is a conserved quantity. 
When a charge {$q$}
 located within the sphere $\manS_\RegI$ passes
 though the throat of the wormhole to $\manM_\RegII$,
 an observer in $\manM_\RegI$
 who has merely integrated $\Hform$ over $\manS^\RegI$
 to establish the conserved quantity ${\Charge}_t^\RegI$,
 no longer sees $q$ in their part of the universe. 
They rather say that after the charge has passed though the throat,
 the wormhole has \textit{gained} charge $q$
 \cite{Visser-LW,Susskind-2005arXiv,Wheeler-1957ap}.

There may still be aspects 
 of this standard ``charged wormhole'' view
 that worry some. 
Of course, 
 if a charge enters a box,
 the charge will still be in the box whenever we subsequently look inside, 
 and we can reasonably say that the box has acquired that charge. 
However,
 in the current case, 
 after passing to universe $\manM_\RegII$, 
 the charge $q$ might subsequently move \textit{arbitrarily far}
 from the throat\footnote{Ignoring dynamical constraints it could
  even move to $\infty_\RegII$ and therefore
  pass out of universe $\manM_\RegII$,
  changing the overall charge of the biverse!}.
At such a distance,
 some might consider it unreasonable to have the 
 steady-state field of the charge
 still influenced by some prehistoric
 transit from $\manM_\RegI$.
Nevertheless,
 the 
 standard viewpoint insists
 that an observer in $\manM_\RegI$
 still sees
 that the wormhole has acquired,
 \textit{and retained},
 charge $q$.

However, 
 since our biverse scenario has a non-trivial topology, 
 we can
 again consider $\Hform$ to be undefined in an absolute sense. 
Having decided that $\Hform$ is not defined,
 one is free to consider how to replace it. 
We consider here a simple extension to Maxwell's equations 
 in which the charge the wormhole gains depends on the distance from
 the charge to the throat, 
 and is \textit{no longer} affected by whether or not
  the charge made a one-way transit
  through that throat in the past.

\newcommand\rthroat{r_{\textup{th}}} 

Consider a single point charge $q$, 
 and define ${{\Charge}^\RegI({r_p})}$ as a function of the radial position,
 ${r_p}$ of the charge. 
Thus we can set
\begin{align}
  {{\Charge}^\RegI({r_p})}
&=
  {q}\,{{\zeta}({r_p})}
\\
  \qquadtext{where}
  {{\zeta}({r_p})}
&=
  \frac12
 +
  \frac{\delta {r_p}}{2(\rthroat^2 + {\delta {r_p}}^2)^{1/2}}
,
\label{Worm_def_zeta}
\end{align}
and where 
 $\rthroat$ is the radius of the throat
 and ${\delta {r_p}}$ is the distance to the throat
 with ${\delta {r_p}}>0$ and ${\delta {r_p}}<0$ 
 if the charge is in $\manM_\RegI$
 and $\manM_\RegII$ respectively.
Although this function is arbitrary it does have the useful feature
 that ${{\Charge}^\RegI(r_p)} \to q$ if ${r_p} \to \infty_\RegI$ and
 the ${{\Charge}^\RegI(r_p)} \to 0$ if ${r_p} \to \infty_\RegII$, 
 and is inline with physical intuition.

We note below that the field $\Hform$
 is still well-defined as long as the charge is moving slowly, 
 $d {r_p}/d t\approx 0$. 
Let $\Fform^\RegI_{r_p}\in\Gamma\Lambda^2 \manM$
 and $\Fform^\RegII_{r_p}\in\Gamma\Lambda^2 \manM$
 be the static electromagnetic field for a point charge at ${r_p}$,
 so that
\begin{align}
  {\de} \Fform^\RegI_{r_p} = 0
\,,\qquad
  {\de} \Fform^\RegII_{r_p} = 0
\,,\qquad
  {\de} {\hstar} \Fform^\RegI_{r_p} =  \Jform_{r_p}
,
\nonumber
\\
\qquad
\qquadand
  {\de} {\hstar} \Fform^\RegII_{r_p} =  \Jform_{r_p}
,
\label{Worm_def_F}
\end{align}
where $\Jform_{r_p}$ is the distributional source
 corresponding to a point charge at 
 $r={r_p}$, 
 $\theta=0$ and $\phi=0$
\begin{align}
  \Jform_{r_p}
=
  \,q{{\delta}(r-r_p)}\,{{\delta}(\theta)}\,{{\delta}(\phi)}\,
  r^2 \sin\theta ~{dr} \wedge {d\theta} \wedge {d\phi}
,
\label{Worm_def_J}
\end{align}
subject to the boundary conditions
\begin{align}
  \lim_{r\to\infty_\RegI} 
  \Fform^\RegI 
=
  0
\,,&
 \qquad
  \lim_{r\to\infty_\RegII} 
  r^2 \Fform^\RegI 
 = 0,
\nonumber
\\
  \lim_{r\to\infty_\RegI} r^2 \Fform^\RegII 
=
  0
  \,,&
  \qquad
  \lim_{r\to\infty_\RegII} \Fform^\RegII 
 =
   0
.
\label{Worm_def_F_bdd_cond}
\end{align}
That is to say the field lines for $\Fform^\RegI$
 due to the point charge all terminate at $\infty_\RegI$,
 whereas those for $\Fform^\RegII$ terminate at $\infty_\RegII$.
Let
\begin{align}
  \Hform_{r_p} 
=
  {{\zeta}({r_p})}\,{\hstar} \Fform^\RegII_{r_p} 
+ 
  \left[1-{{\zeta}({r_p})}\right]
  \,
  {\hstar} 
  \Fform^\RegI_{r_p}
\label{Worm_def_H}
\end{align}
which has the property that,
 as long as the point charge is inside $\manS^\RegI$, which includes
 all of $\manM_\RegII$, then
\begin{align}
  {\Charge}^\RegI_{r_p}
=
  \int_{\manS^\RegI} \Hform_{r_p}
.
\label{Worm_res_H_Q}
\end{align}
Thus as the point charge moves closer to the throat
 more of the field lines reach $\infty_\RegII$,
 fig. \ref{fig_Wormhole_fieldlines}.
However we only approximately solve Maxwell's equations since
\begin{align}
  {\de}\, \Hform_{r_p}
&=
  \Jform_{r_p}
 +
  {{\zeta}'(r_p)}\,\dfrac{r_p}{t}\, {dt} \wedge {\hstar}
  \left(\Fform^\RegI_{r_p}-F^\RegII_{r_p}\right)
\nonumber
\\
&\qquad
 +
  {{\zeta}(r_p)}\,\dfrac{r_p}{t}\, {dt} \wedge {\hstar}
  \left(
    \dfrac{\Fform^\RegI_{r_p}}{r_p} 
   -
    \dfrac{\Fform^\RegII_{r_p}}{r_p}
  \right)
.
\label{Worm_res_dH}
\end{align}
However, we again emphasise that we cannot define a $\Hform_{r_p}$
 which solves both Maxwell's equations \textit{and} eqn. \eqref{Worm_res_H_Q}.

Another attractive feature of our proposed modification occurs in
relation to a wormhole connecting two distinct regions ($\manifold{A}$
and $\manifold{B}$, say) in the same universe.  In this topology, a
charge ${q}$ can circulate multiple ($n$, say) times by entering at
$\manifold{A}$ and exiting at $\manifold{B}$.  Standard Maxwell theory
then predicts that $\manifold{A}$ has a charge of $n{q}$, and
$\manifold{B}$ a charge of $-n{q}$, which can become arbitrarily
large.  The modification to Maxwell's theory of
\eqref{Worm_def_F_bdd_cond} avoids this problem, as integrating
around $\manifold{A}$ will yield a charge that does not exceed $q$.

%
\section{Discussion}
\label{ch_Discussion}

\def\StressEnergyT{\emVec{T}}

\UPDATED{There are mechanisms for charge conservation
 that exist independently of the topology or gauge-free conditions
 that we have discussed above.
One of the most notable is a consequence of Noether's theorem 
 for a $U(1)$ gauge invariant Lagrangian, 
 which enforces local charge conservation ${\de} \Jfree = 0$.
For example,
 if $\Lambda[\mForm{A},\alpha] \in \Gamma \Lambda^4 \manM$
 is invariant under substitutions
 $\alpha \rightarrow e^{\imath \phi} \alpha$
 and
 $\mForm{A} \rightarrow \mForm{A} + \imath {\de}\phi$,
 then the 3-form $\partial\Lambda/\partial\mForm{A}$ 
 is locally conserved,
 i.e. ${\de}( \partial\Lambda/\partial\mForm{A} ) = 0$.
Since the variations are purely local,
 this makes no statement about the \emph{global} conservation of charge
 in non trivial spacetimes. 
It should also be noted that most Lagrangian formulations
 of electromagnetism implicitly assume a model for $\Hform$. 
For example the Maxwell vacuum where $\Lambda$ contains the term 
 $\Lambda^\textup{EM} 
  = \frac{1}{2} {\de}\mForm{A} \wedge \hstar {\de}\mForm{A}$,
 or a model of
 a simple non-dispersive ``antediluvian'' medium\footnote{\UPDATED{A
  a non-dispersive medium would not produce rainbows.}}
 where 
 $\Lambda^\textup{EM} 
  = \frac{1}{2} {\de}\mForm{A} \wedge \hstar \mForm{Z}({\de}\mForm{A})$
 and $\mForm{Z}$ is a constitutive tensor
 \cite{Gratus-T-2011jmp}.
It would also be interesting to attempt to construct Lagrangians
 which do not imply a well defined excitation 2-form.}

\UPDATED{We
 might also broaden our examination of conservation laws
 beyond just charge 
 to those of energy and momentum, 
 by looking at the divergence-free nature
 of the stress-energy tensor $\StressEnergyT$.
In our discussion of section \ref{ch_Singular},
 the total energy of the current and electromagnetic field
 must be zero before the singularity, 
 i.e. on a hypersuface in $\manM^{-}$.
Likewise,
 although we did not define the energy,
 the existence of fields after the singularity,
 implies that the total energy would be non zero. 
However, 
 just as in the case of charge conservation, 
 this lack of global energy conservation is not 
 inconsistent with the local energy conservation law
 ${\de}({\mForm{\tau}_{K}}) = 0$,
 obtained from the energy 3-form $\mForm{\tau}_{K} = \hstar \StressEnergyT({K},-))$,
 where $K$ is a timelike Killing field. 
Of course, 
 in the general relativistic case of an evaporating black hole
 there are challenges about defining the total energy,
 but one should not be surprised if an appropriate measure of total energy 
 were also not conserved.}

\UPDATED{For momentum, 
 if
 $\manM$ possesses a spacelike Killing vector ${K}$,
  then ${K}$ is locally conserved, i.e. ${\de}({\mForm{\tau}_{K}}) = 0$,
 but again this has said nothing about 
 the \emph{global} conservation of momentum. 
We see from \eqref{eg_J+_alt} that the construction of $\Jfree$
 that it is spherically symmetric
 and hence will not change the total momentum. 
However, 
 this was a \emph{choice}
 and non-spherically symmetric currents can easily be 
 obtained by introducing a Lorentz boost. 
Of course, 
 when considering the total momentum, 
 \UPDATED{i.e. that of the electromagnetic field
 plus that of the response of the medium,} 
 one encounters the thorny issue of the Abraham-Minkowski controversy 
 \cite{Dereli-GT-2007pla,Dereli-GT-2007jpa}
 and choice of Poynting vector \cite{Kinsler-FM-2009ejp}.
From the perspective here,
 the question of which momentum is most appropriate
 would be further complicated by the non existence of the excitation 2-form.}

\UPDATED{As a final remark, 
 our results presented here raise the possibility of developing
 a way to prescribe dynamic equations
 for the electromagnetic field $\Fform$
 without introducing or referring to an excitation field $\Hform$ at all. 
One possibility is to combine
 Maxwell's equation \eqref{CQ_Max_HJ}
 directly with the constitutive relations, 
 thus eliminating the need for $\Hform$ \cite{Gratus-MK-AREA51}.}

%
\section{Conclusion}
\label{ch_Conclude}

In this paper we have clarified physical issues 
 regarding electromagnetism on spacetimes with a non-trivial topology --
 either missing points, 
 as 
 can be introduced by
 the singularity at the heart of a black hole, 
 or the presence of wormhole-like bridges between universes,
 or between two locations in the same universe.

We have unambiguously shown that such cases have significant 
 implications for charge conservation --
 i.e. that it need not be conserved; 
 and the role of (or need for)
 the electromagnetic excitation field $\Hform$ 
 (i.e. the Maxwell {\emDH} vector fields) --
 i.e. that it is not always globally unique, 
 and thus has a subordinate or even optional status 
 as compared to the more fundamental $\Fform$ 
 comprising the Maxwell {\emEB} vector fields.
All of these considerations are purely electromagnetic,
 and are prior to any considerations about
 the physics of 
 singularities,
 such as cosmic censorship hypotheses.
\UPDATED{Similar statements
 can be made about the global versus local conservation
 of leptonic and baryonic charges.}

Although our results show that Maxwell's equations
 need not conserve charge on topologically non-trivial spaces,
 neither do they guarantee that they will not (or cannot).
But they do insist that
 charge conservation is not a fundamental property,
 and can only be maintained with additional assumptions.
Further, 
 wormhole mouths do not --
 or need not --
 be considered to accumulate a charge that is the sum of 
 all charge that passes through; 
 it is possible to construct
 a {self-consistent} electromagnetic solution
 where the wormhole only temporarily accommodates a passing charge.


\begin{acknowledgements}

{The JG and PK are grateful for the support provided
by STFC (the Cockcroft Institute ST/G008248/1 and ST/P002056/1) and
EPSRC (the Alpha-X project EP/N028694/1). PK would like to acknowledge
the hospitality of Imperial College London.}
\UPDATED{The authors would like to thank the anonymous referees
 for their useful suggestions.}

\end{acknowledgements}

%

\clearpage

\widetext

\section{Popular summary}\label{S-popular}

\large

\begin{center}
{\huge{Is electromagnetism finished yet?}}\\
~\\
\emph{\small{A popular summary for ``Evaporating black-holes, wormholes, and vacuum polarisation:
                    must they always conserve charge?''}}\\
~\\
{Paul Kinsler}
\end{center}

\setlength{\parskip}{1ex}

~\\
~

You can find interesting things, 
 sometimes, 
 in forgotten corners and disused cellars.
We have looked in the obscure filing cabinets
 holding the planning permissions of Maxwell's explanation
 of electromagnetism
 and found not only a loophole controversial enough 
 to warrant a sign saying \emph{``Beware of the leopard''}, 
 but an equally disconcerting philosophical consequence.
A scientific paper
 published in 
 Foundations of Physics\footnote{See
     {{\lq\lq}Evaporating black-holes, wormholes,
     vacuum polarisation: must they always conserve charge?{\rq\rq}},\\
      J. Gratus, P. Kinsler, M.W. McCall, \qquad
        {Found. Phys. \textbf{49}, 330 (2019)},\\
      {http://doi.org/10.1007/s10701-019-00251-5}, \qquad
        {https://arxiv.org/abs/1904.04103}.}
, 
 explains the mathematical basis for the loophole in forensic detail.
In what follows 
 I present a simplified description of this work 
 by 
 Jonathan Gratus, myself, and Martin McCall.

To explain this we start with
 the mathematical equations for electromagnetism
 that were famously first developed by James Clerk Maxwell.
Equations and mathematical models are key
 to the process and predictions of physics, 
 since they turn our understanding into a tool
 for getting precise answers.
Maxwell's equations tell us about 
 how electric charges make electric fields, 
 how electric currents make magnetic fields;
 and even how combinations of oscillating electric and magnetic fields
 can form electromagnetic waves.
These electromagnetic waves make up the 
 the light we see with,
 as well as the radio waves, 
 mobile phone signals,  
 X-rays, 
 and so on.
They provide a great deal of 
 the communications and imaging technology
 that we use everyday in our modern civilization.

Here, 
 we just need to know that there are not two but \emph{four} fields
 in Maxwell's equations.
In addition to the well-known electric and magnetic fields, 
 there is also a pair of ``excitation'' fields
 which act as a complement to the ordinary fields.
%
It is in fact
 these excitation fields that are directly linked 
 to charge and current by Maxwell's equations, 
 and not the more famous electric and magnetic fields.
Furthermore, 
 the sources of these fields -- 
 electrical charges and currents --
 are always seen to be conserved:  
 they do not suddenly appear from nowhere or disappear to nowhere, 
 and we can track them if they move around.
Conservation of charge is widely believed
 to follow as a consquence
 of Maxwell's equations themselves, 
 although there are also other justifications.

The effect of the loophole we have found is this:
 even if charge is perfectly conserved at every individual place and time, 
 the total charge of the universe might change anyway.
However, 
 to expose the loophole in such a well-established structure
 as electromagnetism
 is not easy.
To show the potential for a failure for global charge conservation,
 you have to walk a very careful line, 
 one that requires both an exotic spacetime
 and a fresh look at the electromagnetic excitation fields.
Whilst the electric field and magnetic field, 
 familiar to all secondary school science pupils,
 are an uncontestable part of the structure of electromagnetism,
 the status of their excitation field counterparts
 is less robust.

In an ordinary run-of-the-mill universe, 
 with a ``topologically trival'' spacetime that lacks any holes, 
 Maxwell's equations alone can guarantee charge conservation.
Topologically trival spacetimes 
 make it easy to do your sums
 and show that global charge conservation
 always follows from local conservation;
 but the presence of a hole
 will break that simplicity and makes the answers uncertain.
Next, 
 if the excitation fields are real and physical,
 Maxwell's equations also can guarantee charge conservation
 without help.
But if the spacetime has a hole in it, 
 \emph{and}
 the excitation fields do not have a real physical existence, 
 Maxwell theory by itself becomes unable to assert  
 a cast-iron law of charge conservation.

So what sort of spacetime has the non-trivial topology
 required to incite otherwise mild-mannered physicists
 to try breaking one of the almost sacred tenets of electromagnetism?
We can present two possibilities --
 firstly, 
 a universe in which a black hole forms and then evaporates, 
 so its central singularity 
 creates the necessary temporary hole in spacetime; 
 and
 secondly, 
 where different regions of an otherwise unremarkable universe
 are connected by a wormhole.
In such spacetimes, 
 if the excitation fields are only aides to calculation, 
 and might not have unique values,
 then proofs of global charge conservation suddenly develop an Achilles heel.

Firstly, 
 in the black hole case,
 events can occur where conservation-breaking fields and charge appear
 around its central singularity --
 even though that singularity does not exist within the universe's spacetime, 
 so that there is no place the charge could have come from.
A charge emerging in this way is not only
 an event \emph{without} a cause, 
 but an event for which there is literally nowhere
 for any imagined cause to have existed.
And
 in a twist designed to baffle philosophers, 
 if a patch had been applied to the universe
 in order to provide somewhere for the rogue charge
 to have actually been created,
 we would suddenly find that
 the charge non-conservation we were trying to explain
 would now be utterly forbidden!

Secondly, 
 this analysis also means we have to re-evaluate electromagnetism
 in a universe with a wormhole connecting two locations, 
 albeit with consequences that are 
 hopefully less likely to induce anxiety amongst physicists and philosophers.
Here, 
 the traditional view says that if a charged object
 were to pass through the wormhole, 
 the lines of electric field 
 that spread out from the charge
 are dragged behind it like a very long tail.
Then, 
 because the wormhole's exit now has a bundle of field lines
 (a `tail')
 heading back into it, 
 that exit has an effective charge opposite to the object; 
 likewise,  
 the wormhole's entrance has the field lines emerging in a spray, 
 just as if it were charged itself.
The wormhole has, 
 in effect, 
 become charged, 
 and this effect must persist no matter
 how far the charge moves away from the wormhole, 
 and no matter how long ago it happened.
Remarkably, 
 if it is possible to loop around and back through the wormhole entrance
 in the same direction, 
 again and again, 
 the loops of field line ``tail'' keep building up,
 as if you were darning a sock,
 leaving the wormhole ever more strongly charged.
This traditional argument insists that 
 no matter how long you wait, 
 the field pattern of a charge therefore can always depend, 
 not only on the position of the charge and the shape of the universe.
It can --
 it \emph{must} --
 {also} depend
 on any transits through wormholes, 
 however far back in prehistory they might have occurred.

In contrast, 
 the loophole we have discovered 
 means that the traditional view no longer holds -- 
 the field pattern, 
 if left long enough, 
 can forget those past wormhole transits.
The field-line tail still exists, 
 but it is no longer wound around in loop after loop through the wormhole.
The interesting topology, 
 and the non-physical nature of the excitation fields
 means that the charging of wormholes need not be permanent --
 the field lines now need only depend 
 on the position of the charge and the shape of the universe.

The results therefore presents us with a challenge -- 
 not only to charge conservation, 
 but to how Maxwell's theory of electromagnetism is interpreted.
It defies us one hand with an effect-without-a-cause philosophical conundrum,
 but on the other conveniently allows wormholes to forget
 their possibly chequered past assisting charged particles
 to take shortcuts through spacetime.
Such controversial findings suggest that our understanding of
 electromagnetism might still have some way to go.

\end{document}